\documentclass[manuscript]{aastex}
\usepackage{amsmath}
\usepackage{arydshln}
\usepackage{multirow}
\usepackage{lscape}

\usepackage{grffile}

\newcommand{\cs}{$J=5-4$}
\newcommand{\so}{$6_7-5_6$}
\newcommand{\SO}{$^{34}$SO}
\newcommand{\tfso}{$6_5-5_4$}
\newcommand{\SOO}{SO$_2$}
\newcommand{\soo}{$10_{3, 7}-10_{2, 8}$} 
\newcommand{\sio}{$J=6-5$}

\newcommand{\alllines}{CS, SO, \SO, \SOO, and SiO}
\newcommand{\linesexceptCS}{SO, \SO, \SOO, and SiO}

\newcommand{\HCO}{HCO$^+$}
\newcommand{\FA}{NH$_2$CHO} 
\newcommand{\MN}{CH$_3$OH}
\newcommand{\MF}{HCOOCH$_3$}
\newcommand{\DME}{CH$_3$OCH$_3$}
\newcommand{\cycCH}{c-C$_3$H$_2$} 
\newcommand{\MT}{CH$_4$} 
\newcommand{\FAD}{H$_2$CO}
\newcommand{\TFA}{H$_2$CS}
\newcommand{\Water}{H$_2$O}
\newcommand{\COO}{CO$_2$}

\newcommand{\mf}{$20_{5, 16}$--$19_{5, 15}$; E}
\newcommand{\dme}{$13_{5, 8}$--$13_{4, 9}$; EE, AA}
\newcommand{\cycch}{$5_{2, 3}$--$4_{3, 2}$} 
\newcommand{\cch}{$N$=3--2, $J$=5/2--3/2, $F$=3--2, 2--1} 

\newcommand{\ALMA}{Atacama Large Millimeter/submillimeter Array (ALMA)}
\newcommand{\iras}{IRAS 16293--2422}

\newcommand{\ire}{infalling-rotating envelope}

\newcommand{\cb}{centrifugal barrier} 

\newcommand{\desys}{disk/envelope system}
\newcommand{\sysV}{systemic velocity}
\newcommand{\ia}{inclination angle}

\newcommand{\bolL}{bolometric luminosity}
\newcommand{\bolT}{bolometric temperature}

\newcommand{\desT}{desorption temperature}
\newcommand{\desE}{desorption energy}

\newcommand{\Ed}{$E_{\rm des}$}
\newcommand{\Td}{$T_{\rm des}$}
\newcommand{\Msun}{$M_\odot$}

\newcommand{\Lsun}{$L_\odot$}

\newcommand{\rcb}{$r_{\rm CB}$}
\newcommand{\rCB}{\rcb}

\newcommand{\vsys}{$v_{\rm sys}$}

\newcommand{\vfall}{$v_{\rm fall}$}
\newcommand{\vrot}{$v_{\rm rot}$}

\newcommand{\molratio}[2]{$N$({#1})/$N$({#2})}

\newcommand{\inv}{$^{-1}$}
\newcommand{\kmps}{km s\inv}
\newcommand{\Jypb}{Jy beam\inv}
\newcommand{\fdegr}{.\!\!\degr}
\newcommand{\hydro}{H$_2$}
\newcommand{\nhydro}{$n$(\hydro)}
\newcommand{\coldenshydro}{$N$(\hydro)}
\newcommand{\cmcubic}{cm$^{-3}$}
\newcommand{\cmsquare}{cm$^{-2}$}

\newcommand{\vsysval}{4.0}

\newcommand{\incRem}{0\degr\ for a face-on configuration}

\newcommand{\gtsim}{\protect\raisebox{-0.7ex}{$\:\stackrel{\textstyle >}{\sim}\:$}}

\newcounter{tbnotecount}
\newcommand{\tbnote}{\refstepcounter{tbnotecount}\alph{tbnotecount}}
\newcommand{\tbnotemark}{\tablenotemark{\tbnote}}
\newcommand{\tbnotetext}[1]{\tablenotetext{\tbnote}{#1}}
\newcommand{\steptbnote}{\refstepcounter{tbnotecount}}
\newcommand{\resettbnote}[1]{\setcounter{tbnotecount}{0}#1}

\newcommand{\iffigure}{\iftrue}

\newcommand{\bfresubfirst}{}
\newcommand{\bfresubsecond}{}
\renewcommand{\bf}{}

\shorttitle{Elias 29 Disk}
\shortauthors{Oya et al.}

\title{Sulfur-Bearing Species Tracing the Disk/Envelope System in the Class I Protostellar Source Elias 29} 

\author{Yoko Oya\altaffilmark{1}, 
Ana L{\'o}pez-Sepulcre\altaffilmark{2, 3, 4}, 
Nami Sakai\altaffilmark{5}, 
Yoshimasa Watanabe\altaffilmark{6, 7. 8}, \\
Aya E. Higuchi\altaffilmark{5}, 
Tomoya Hirota\altaffilmark{9}, 
Yuri Aikawa\altaffilmark{10}, 
Takeshi Sakai\altaffilmark{11}, 
Cecilia Ceccarelli\altaffilmark{3, 4}, \\
Bertrand Lefloch\altaffilmark{3, 4}, 
Emmanuel Caux\altaffilmark{12, 13}, 
Charlotte Vastel\altaffilmark{12, 13}, 
Claudine Kahane\altaffilmark{3, 4}, \\
and Satoshi Yamamoto\altaffilmark{1, 14}} 

\email{oya@taurus.phys.s.u-tokyo.ac.jp}
\altaffiltext{1}{Department of Physics, The University of Tokyo, 7-3-1, Hongo, Bunkyo-ku, Tokyo 113-0033, Japan}
\altaffiltext{2}{Institut de Radioastronomie Millim\'etrique (IRAM), 38406, Saint Martin d'H\'eres, France}
\altaffiltext{3}{Universit\'e Grenoble Alpes, IPAG, F-38000 Grenoble, France}
\altaffiltext{4}{Le Centre National de la Recherche Scientifique (CNRS), IPAG, F-38000 Grenoble, France}
\altaffiltext{5}{RIKEN Cluster for Pioneering Research, 2-1, Hirosawa, Wako-shi, Saitama 351-0198, Japan}
\altaffiltext{6}{Division of Physics, Faculty of Pure and Applied Sciences, University of Tsukuba, Tsukuba, Ibaraki 305-8571, Japan}
\altaffiltext{7}{Tomonaga Center for the History of the Universe, Faculty of Pure and Applied Sciences, University of Tsukuba, Tsukuba, Ibaraki 305-8571, Japan}
\altaffiltext{8}{College of Engineering, Nihon University, 1 Nakagawara, Tokusada, Tamuramachi, Koriyama, Fukushima 963-8642, Japan}
\altaffiltext{9}{National Astronomical Observatory of Japan, Osawa, Mitaka, Tokyo 181-8588, Japan}
\altaffiltext{10}{Department of Astronomy, The University of Tokyo, 7-3-1, Hongo, Bunkyo-ku, Tokyo 113-0033, Japan} 
\altaffiltext{11}{Department of Communication Engineering and Informatics, Graduate School of Informatics and Engineering, The University of Electro-Communications, Chofugaoka, Chofu, Tokyo 182-8585, Japan}
\altaffiltext{12}{Universite de Toulouse, UPS-OMP, F-31028 Toulouse Cedex 4, France}
\altaffiltext{13}{Le Centre National de la Recherche Scientifique (CNRS), IRAP, 9 Av. Colonel Roche, BP 44346, F-31028 Toulouse Cedex 4, France}
\altaffiltext{14}{Research Center for the Early Universe, The University of Tokyo, 7-3-1, Hongo, Bunkyo-ku, Tokyo 113-0033, Japan}

\begin{abstract}
We have observed the Class I 
protostellar source Elias 29 with Atacama Large Millimeter/submillimeter Array (ALMA). 
We have detected CS, SO, $^{34}$SO, SO$_2$, and SiO line emissions 
in a compact component concentrated near the protostar 
and a ridge component separated from the protostar by 4\arcsec\ ($\sim 500$ au). 
The former component is found to be abundant in SO and SO$_2$ 
but
deficient in CS. 
The abundance ratio SO/CS is as high as 
$3^{+13}_{-2} \times 10^2$ at the protostar, 
which is 
even higher than that in the outflow-shocked region of L1157 B1. 
However, 
organic molecules (HCOOCH$_3$, CH$_3$OCH$_3$, CCH, and c-C$_3$H$_2$) 
are deficient in Elias 29. 
We attribute the deficiency in organic molecules 
and richness in SO and SO$_2$ to 
the evolved nature of the source 
or the relatively high dust temperature 
(\protect\raisebox{-0.7ex}{$\:\stackrel{\textstyle >}{\sim}\:$} 20 K) 
in the parent cloud of Elias 29. 
The SO and SO$_2$ emissions trace  
rotation around the protostar. 
Assuming a {\bf highly inclined configuration ($i \geq 65$\degr; 0\degr\ for a face-on configuration)} and Keplerian motion for simplicity, 
the protostellar mass is estimated to be (0.8 -- 1.0) \Msun. 
The $^{34}$SO and SO$_2$ emissions are asymmetric in their spectra; 
the blue-shifted components are weaker than the red-shifted ones. 
Although 
this may be attributed to the asymmetric molecular distribution, 
other possibilities are also discussed.

\end{abstract}


\begin{document}
\section{Introduction} \label{sec:intro}

{\bfresubfirst Because} star formation is a gravitational collapse of a parent molecular cloud core, 
{\bfresubfirst the} 
physical and chemical evolution 
{\bfresubfirst of protostellar sources} 
is affected by environmental effects on the {\bfresubfirst parent} core. 
In addition to the physical diversity of newly born protostars, 
{\bfresubfirst such as} in a single, multiple, or cluster form, 
{\bfresubfirst the} chemical diversity of protostellar sources has recently been recognized {\bfresubfirst on} 
{\bfresubsecond size scales} from 1000 au {\bfresubfirst down} to 10 au 
\citep{Sakai_carbonchain, Lindberg2012, Sakai_ChemRev, Graninger2016, Imai2016, Oya_16293, Oya_483, Lindberg2016, LeeCF2017, Higuchi2018, Lefloch2018}. 
{\bfresubfirst The} 
relation 
{\bfresubfirst between physical and chemical diversities and} 
environment 
has attracted broad attention in astrochemistry, 
astrophysics, and planetary science. 
To assess this problem, it is important to study physical and chemical structures 
of protostars in various evolutionary stages and environments.

Elias 29 (WL 15) is a 
Class I protostar in the L1688 dark cloud in Ophiuchus 
\citep{Elias_rhoOphiuchi, WilkingLada_rhoOphiuchi}, 
whose distance is 137 pc \citep{Ortiz-Leon_OphiuchusDist}. 
The \bolT\ and \bolL\ are reported to be 
{\bfresubfirst 391 K and 13.6 \Lsun\ \citep{Miotello2014},} 
respectively. 
This source is surrounded by a number of young stellar objects, 
{\bfresubfirst such as} WL 16, WL 17, WL19, and WL20. 
Moreover, its parent cloud (L1688) is strongly illuminated by the B2 V star HD147889 
and is a typical photodissociation region 
\citep{Yui1993, Liseau1999, Ebisawa2015, RochaPilling2018}. 
Specifically, Elias 29 is located at $\sim 700$\arcsec\ ($\sim 0.5$ pc) 
from HD147889 in the plane of the sky.

{\bfresubfirst Because} Elias 29 is a bright {\bfresubfirst infrared} source, 
infrared spectroscopic observations have been conducted {\bfresubfirst on its gas and dust components.} 
\citet{Boogert2000} {\bfresubfirst observed Elias 29 with the Infrared Space Observatory (ISO) and found} that CO mainly exists {\bfresubfirst as a} gas, 
while CO$_2$ {\bfresubfirst is} in the solid phase. 
They concluded 
that gas and dust around the protostar 
are significantly heated by external/internal radiation. 
Recently, \citet{RochaPilling2018} showed, on the basis of a radiative transfer calculation, 
that the dust temperature of this protostellar core is mostly higher than 20 K 
{\bfresubfirst owing} to external irradiation, especially {\bfresubfirst from} HD147889. 
{\bfresubfirst Because} the \desT\ of CO is about 20 K, 
their result is consistent with the ISO observation. 
Elias 29 is also an X-ray {\bfresubfirst emitter} 
\citep{Imanishi2001, Favata2005, Giardino2007}. 
The effect of {\bfresubfirst high-energy cosmic rays} 
on the chemical composition of gas and dust {\bfresubsecond of Elias 29} 
is discussed by \citet{RochaPilling2015}. 

Molecular outflows from Elias 29 have extensively been studied by 
\citet{Bontemps1996}, \citet{Sekimoto1997}, \citet{Ceccarelli2002}, \citet{Bussmann2007}, 
\citet{Nakamura2011}, and \citet{vanderMarel2013}. 
According to the CO ($J=6-5$) observation {\bfresubfirst performed} by \citet{Ceccarelli2002} 
at a resolution of 12\arcsec\ with {\bfresubfirst the} James Clark Maxwell Telescope (JCMT), 
the outflow of Elias 29 {\bfresubfirst is} along the east-west direction, 
where the eastern and western lobes are red-shifted and blue-shifted, respectively. 
\citet{Bussmann2007} conducted the CO ($J=3-2$) observation with 
{\bfresubfirst the} Heinrich Hertz Submillimeter Telescope (HHT) 
and {\bfresubfirst found} that the outflow of this source has an inverse S shape 
{\bfresubfirst that} 
{\bfresubsecond lies} along 
the east-west direction near the protostar 
and along 
{\bfresubfirst the north-south direction at a distance of 10000 au} 
from 
{\bfresubsecond the protostar.} 
{\bfresubfirst The direction of the outflow near the protostar} 
is consistent with the {\bfresubfirst observations of} \citet{Ceccarelli2002}. 
Meanwhile, \citet{Bussmann2007} and \citet{Nakamura2011} showed that 
the large-scale outflow {\bfresubfirst of Elias 29} 
is complex owing to 
{\bfresubfirst outflow} contributions of nearby young stellar objects. 
\citet{vanderMarel2013} conducted {\bfresubsecond a} CO ($J=3-2$) observation with {\bfresubfirst the} JCMT 
and found {\bfresubfirst an} outflow shape 
consistent with 
that reported by \citet{Bussmann2007}. 
In addition to the molecular outflow, 
a jet launched from the protostar toward the 
{\bfresubfirst east-west} 
{\bfresubsecond direction} 
was {\bfresubfirst detected in} near-infrared \hydro\ emission \citep{Gomez2003, Ybarra2006}.

The distribution of the dense gas around the protostar is delineated by 
\citet{Boogert2002_Elias29}, \citet{Lommen2008}, \citet{Jorgensen2009}, and \citet{vanKempen2009}. 
With {\bfresubfirst the} Submillimeter Array (SMA), 
\citet{Lommen2008} {\bfresubfirst found} 
{\bfresubfirst a} compact component associated 
{\bfresubfirst with} the protostar 
{\bfresubfirst in} \HCO\ ($J=3-2$) emission at a resolution of $4\farcs0 \times 2\farcs3$. 
{\bfresubfirst Their observation also revealed} 
a ridge component extending along the east-west direction 
{\bfresubfirst at} 4\arcsec\ ($\sim 500$ au) south {\bfresubfirst of} the protostar. 
They estimated the protostellar mass to be $2.5 \pm 0.6$ \Msun\ 
by assuming Keplerian rotation, 
where the inclination of the {\bfresubfirst \desys\ was} assumed to be 30\degr\ (\incRem). 
However, the disk/envelope structure has not been well characterized 
{\bfresubfirst because of} poor angular resolution and sensitivity. 
Moreover, the chemical composition in the protostar {\bfresubfirst vicinity} has not been {\bfresubfirst investigated} yet. 

Here we report the physical and chemical structures of the \desys\ at subarcsecond resolution with {\bfresubfirst the \ALMA.} 
This work is a part of our comparative study of five {\bfresubsecond young} low-mass 
protostellar sources 
{\bfresubfirst (TMC--1A, B335, NGC1333 IRAS 4A, L483, and Elias 29)} 
\citep[][and this work]{Sakai_TMC1A, Imai2016, LopezSepulcre2017, Oya_483}.

\section{Observations} \label{sec:obs}
The ALMA observations of Elias 29 were carried out on 2015 May 18  
with 37 antennas during the Cycle 2 operation. 
Spectral lines of \alllines\ were observed with the Band 6 receiver, 
{\bfresubsecond and the basic parameters of the observations} 
are listed in Table \ref{tb:lines}. 
The {\bfresubsecond baselines} ranged from 15.1 m to 519.8 m. 
The field center of the observations was $(\alpha_{2000}, \delta_{2000}) = (16^{\rm h}27^{\rm m}09\fs44, -24\degr37^\prime19\farcs99)$, 
and the primary beam size (FWHM) {\bfresubfirst was} 23\farcs03. 
The total on-source time was 22.27 minutes. 
The typical system temperature was from 70 to 120 K. 
Sixteen {\bfresubfirst spectral windows} were observed with 
a backend correlator tuned to a resolution of 61.035 kHz (0.073 \kmps\ at 250 GHz), 
and the bandwidth of each {\bfresubfirst window} was 58.5938 MHz. 
J1517--2422 was used for the bandpass calibration, 
while J1625--2527 was used for the phase calibration every 7 minutes. 
An absolute flux density scale was derived from Titan. 
{\bfresubfirst The absolute} accuracy of the flux calibration {\bfresubfirst was} 10\%\ for Band 6 \citep{ALMA-TH2}. 
Self-calibration was not applied, for simplicity. 

Images were obtained {\bfresubfirst with} the CLEAN algorithm 
{\bfresubfirst using} {\bfresubsecond Briggs} weighting with {\bfresubfirst a} {\bfresubsecond robust} parameter of 0.5. 
{\bfresubfirst A} 1.2 mm continuum image was obtained by averaging line-free channels. 
The line maps were obtained after subtracting the continuum component directly from the visibility data 
by resampling to make the channel width 0.5 \kmps\ for \SO\ and SiO or 0.1 \kmps\ for the {\bfresubsecond other molecular transitions.}  
{\bfresubfirst The synthesized} beam sizes for the spectral lines are listed in Table \ref{tb:lines}. 
The root-mean-square (rms) noise level {\bfresubfirst was} 0.3 m\Jypb\ for the continuum 
{\bfresubfirst and} 7, 7, 4, 6, and 4 m\Jypb\ for \alllines, respectively, 
for the channel width mentioned above. 
{\bfresubfirst A} primary beam correction {\bfresubfirst was} applied for the continuum and line maps (see Figure \ref{fig:dist}(a)).

\section{Results: Spectral Distribution} \label{sec:dist}

\subsection{1.2 mm Continuum Emission} \label{sec:dist_cont}

Figure \ref{fig:dist}(a) shows the continuum emission at 1.2 mm. 
{\bfresubsecond {\bf The storing} beam is $0\farcs856 \times 0\farcs470$ (P.A. 95\fdegr17).} 
The peak position of the continuum emission was determined {\bfresubfirst to be} 
$(\alpha_{2000}, \delta_{2000}) = (16^{\rm h}27^{\rm m}09\fs4358\pm0.0007, -24\degr37^\prime19\farcs286\pm0.004)$, 
by using two-dimensional Gaussian fitting. 
The image component {\bfresubfirst sizes} convolved with and deconvolved from the beam are  
$(0\farcs8845\pm0\farcs0216) \times (0\farcs5632\pm0\farcs0092)$ ($\sim 120 \times 80$ au) {\bfresubfirst (Position Angle (P.A.)} $96\fdegr0\pm1\fdegr4$) 
and 
$(0\farcs312\pm0\farcs033) \times (0\farcs220\pm0\farcs134)$ ($\sim 40 \times 30$ au) (P.A. $177\degr\pm57\degr$), 
respectively. 
Thus, the continuum emission is 
marginally resolved by {\bfresubsecond the} {\bf storing} beam. 


{\bfresubfirst Two-dimensional Gaussian fitting of the continuum emission yields a} peak intensity $I(\nu)$ and integrated flux $F(\nu)$ of 
$(17.2\pm0.3)$ m\Jypb\ and $(21.4\pm0.7)$ {\bfresubfirst mJy, respectively,} where the errors represent three standard deviations (3$\sigma$) {\bfresubfirst of} the fit. 
The beam-averaged column density of \hydro\ (\coldenshydro) is {\bfresubfirst obtained} 
{\bfresubfirst from} the following equation \citep{Ward-Thompson_gasmass}: 
\begin{equation}
	N ({\rm H}_2) = \dfrac{2 \ln 2 \cdot c^2}{\pi h \kappa_\nu m R_{\rm d}} 
		\times \dfrac{F(\nu)}{\nu^3 \theta_{\rm major} \theta_{\rm minor}} 
		\times \left(\exp \left(\dfrac{h\nu}{kT} \right) - 1\right), 
\end{equation}
where 
$\kappa_\nu$ is the mass absorption coefficient with respect to the dust mass, 
$m$ is the average mass of a particle in the gas ($3.83 \times 10^{-24}$ g), 
$\nu$ is the frequency, 
$\theta_{\rm major}$ and $\theta_{\rm minor}$ are the major and minor beam sizes, respectively, 
$T$ is the dust temperature, 
and $R_{\rm d}$ is the dust-to-gas mass ratio (0.01). 
According to \citet{Ossenkopf1994}, 
$\kappa_\nu$
is estimated to be
1.3 cm$^2$ g\inv\ at {\bfresubfirst a} wavelength of 1.2 mm by interpolation. 
{\bfresubfirst In this study, 
we chose the dust opacity model appropriate for dense regions ($10^8$ \cmcubic ) with the MRN grain size distributions 
\citep{Mathis1977}. 
{\bfresubsecond The resulting} \coldenshydro\ and gas mass {\bfresubsecond are} 
shown in Table \ref{tb:continuum}. 
Because} Elias 29 is a relatively evolved source, 
the dust mass opacity {\bfresubfirst at 1.2 mm} could be {\bfresubfirst higher 
than the typical value for young stellar objects 
owing to} a larger dust size. 
If we assume $\beta =$ 1.0 \citep{Miotello2014}, 
$\kappa_\nu$ is estimated to be $2.5$ cm$^2$ g\inv\ 
\citep[$\kappa_\nu \times R_{\rm d} = 0.1 \times \left(\nu / (10^{12} {\rm \ Hz}) \right)^\beta$;][]{Beckwith1990}, 
and \coldenshydro\ and 
{\bfresubsecond gas mass would be smaller by a factor of 2.} 
In the following discussion, 
we {\bfresubfirst use} the former results with $\kappa_\nu =$ 1.3 cm$^2$ g\inv.

The peak intensity (17.2 m\Jypb) 
corresponds to {\bfresubfirst a} brightness temperature of 0.8 K with the {\bf storing} beam size. 
When we {\bfresubfirst use} the image size deconvolved from the beam to compensate {\bfresubfirst for} beam dilution, 
the brightness temperature is 4.7 K. 
\citet{RochaPilling2018} reported the distribution of the dust temperature. 
According to their {\bfresubfirst model,} 
the dust temperature is about 70 K at {\bfresubfirst a} distance of 40 au from the protostar. 
With these values, the beam-averaged optical depth of the dust continuum is 
estimated to be 0.07.

\subsection{Molecular Lines} \label{sec:dist_mol}

\begin{landscape}
\begin{figure}
	\iffigure
	\vspace*{-20pt}
	\includegraphics[bb = 0 0 800 450, scale = 0.7]{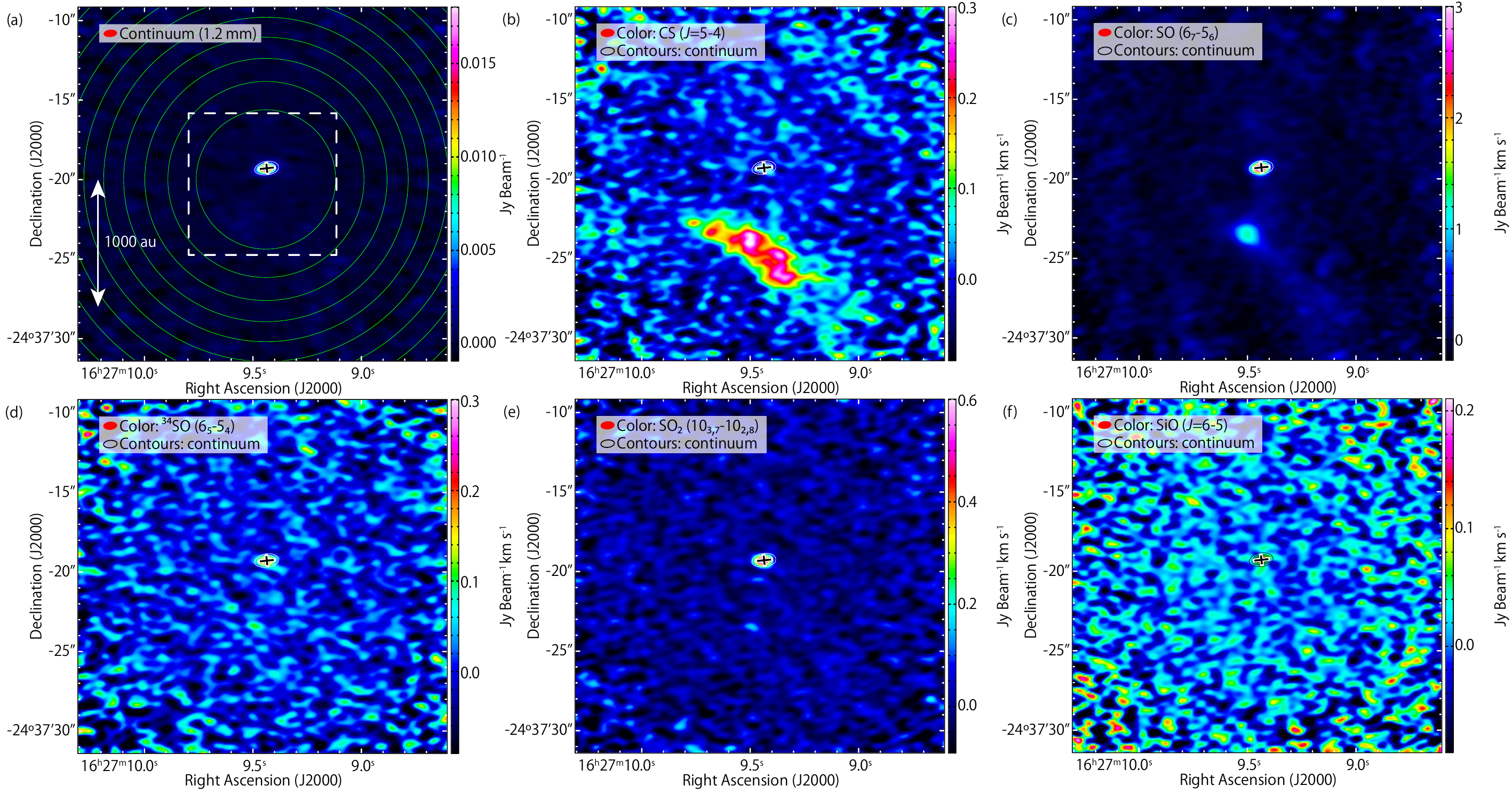}
	\vspace*{-20pt}
	\fi
	\caption{(a) Dust continuum emission at 1.2 mm (254.6 GHz).  
			Integrated intensity maps of 
			the (b) CS (\cs), (c) SO (\so), (d) \SO\ (\tfso), (e) \SOO\ (\soo), and (f) SiO (\sio) {\bfresubfirst emissions.} 
			The velocity range for the integration is $-20$ to 30 \kmps. 
			White and black contours in each panel represent the continuum map, 
			and the contour levels are {\bfresubfirst at intervals of 10$\sigma$ starting} from 10$\sigma$, 
			where the rms level is 0.3 m\Jypb. 
			{\bfresubfirst The} black cross in each panel represents 
			the continuum peak position, 
			and {\bfresubfirst its size represents} the FWHM widths of the image component size convolved with the beam, 
			obtained {\bfresubfirst from} two-dimensional Gaussian fitting. 
			The velocity range for integration is $-20$ to $+30$ \kmps. 
			The images are corrected {\bfresubfirst with} the primary beam 
			represented by the green contours in panel (a), 
			whose contour levels are {\bfresubfirst at intervals of 10\% starting} from 10\% of its peak value. 
			{\bfresubfirst The white} rectangle in panel (a) represents 
			the area for the blowup maps in Figures \ref{fig:dist_blowup} and \ref{fig:mom0_Cbearing}. 
			\label{fig:dist}}
\end{figure}
\end{landscape}

\begin{landscape}
\begin{figure}
	\iffigure
	\includegraphics[bb = 0 0 800 450, scale = 0.7]{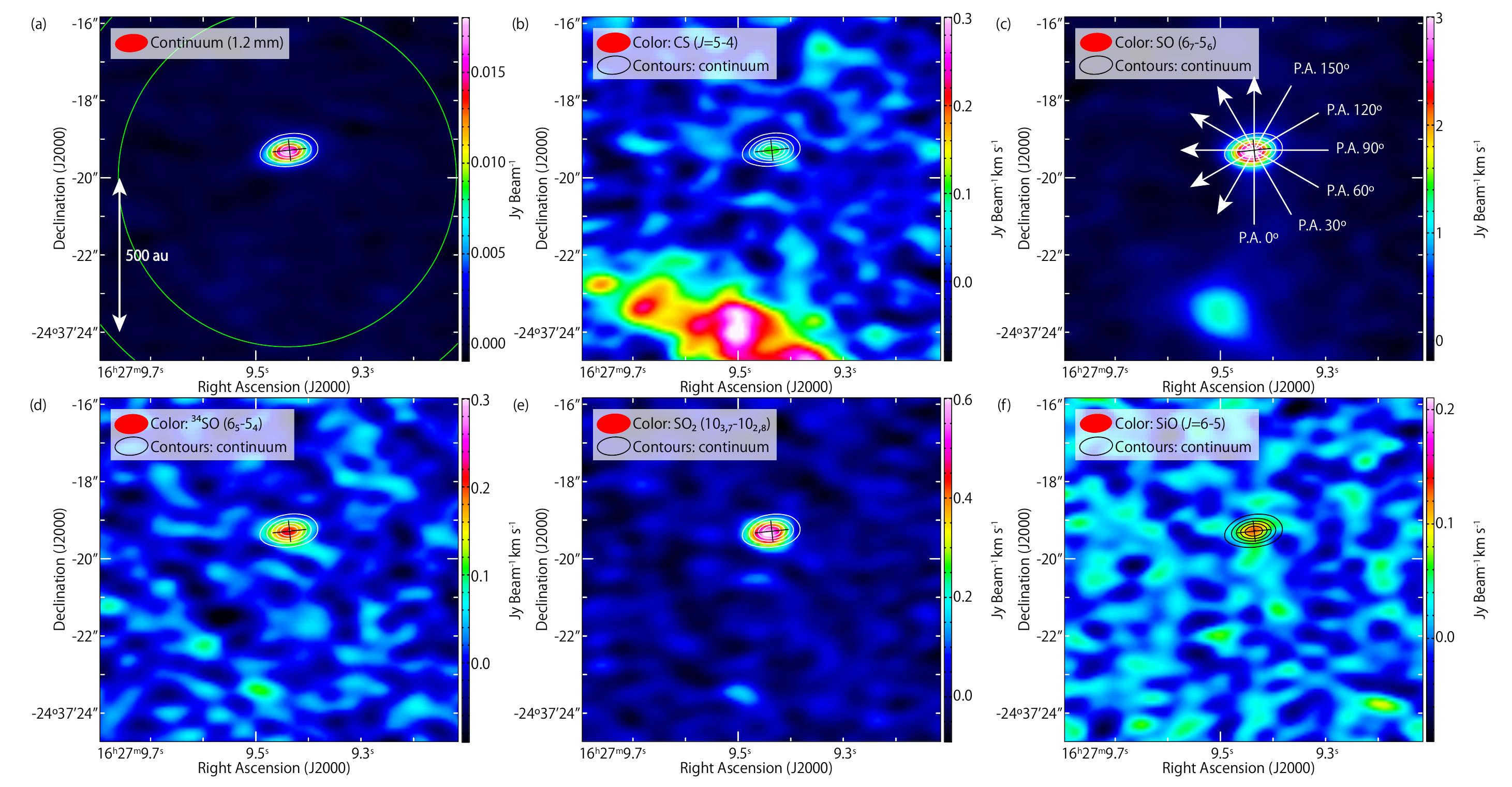}
	\vspace*{-20pt}
	\fi
	\caption{Blowups of Figure \ref{fig:dist}. 
			(a) Dust continuum emission at 1.2 mm. 
			{\bfresubfirst Integrated} intensity maps of the (b) CS, (c) SO, (d) \SO, (e) \SOO, and (f) SiO {\bfresubfirst emissions.} 
			The area for the blowup is represented by the white dashed rectangle in Figure \ref{fig:dist}(a). 
			See the caption of Figure \ref{fig:dist} for details. 
			\label{fig:dist_blowup}}
\end{figure}
\end{landscape}

Figure \ref{fig:dist} {\bfresubfirst shows} the integrated intensity maps of 
the \alllines\ lines. 
The velocity range for the integration is $-20$ to 30 \kmps, 
where the systemic velocity (\vsys) is \vsysval\ \kmps. 
Figure \ref{fig:dist_blowup} shows the blowup of each panel in Figure \ref{fig:dist} around the continuum peak, 
whose area is indicated by the white dashed rectangle in Figure \ref{fig:dist}(a). 
Figure \ref{fig:spectra} shows 
{\bfresubsecond spectrum of each transition} around 
the continuum peak. 
{\bfresubfirst The peak integrated intensities of the molecular emission are obtained 
by using two-dimensional Gaussian fitting 
to their integrated intensity maps (Table \ref{tb:abundance}).} 

\subsubsection{SO, \SOO, and SiO} \label{sec:dist_mol_SO}

The \linesexceptCS\ emissions show 
{\bfresubfirst a point source at the continuum peak position. 
Two-dimensional Gaussian fitting shows the} position of the {\bfresubfirst SO} intensity peak {\bfresubfirst to be} 
$(\alpha_{2000}, \delta_{2000}) = (16^{\rm h}27^{\rm m}09\fs4379\pm0.0008, -24\degr37^\prime19\farcs30\pm0.01)$. 
The image component {\bfresubfirst sizes} (FWHM) of the SO emission 
convolved with and deconvolved from the beam are 
$\sim 1\farcs0 \times 0\farcs7\ (\sim 120\ {\rm au} \times 90\ {\rm au})$ (P.A. $121\fdegr5 \pm 5\fdegr5$) 
and 
$\sim 0\farcs7 \times 0\farcs3\ (\sim 90\ {\rm au} \times 40\ {\rm au})$ (P.A. $158\degr \pm 11\degr$), 
respectively. 

{\bfresubfirst The intensity ratio $^{32}$SO/\SO\ is found to be $13\pm4$ 
from the peak integrated intensities (Table \ref{tb:abundance}). 
Because} the $^{32}$S/$^{34}$S ratio in the Solar neighborhood is about 22.6, 
the $^{32}$SO line is likely optically thick. 
The optical depth of $^{32}$SO is indeed estimated to be 2.33, 
including the correction for the different $S \mu^2$ {\bfresubfirst values} of the two lines. 
Here, we neglect the small difference {\bfresubfirst in} their upper-state energies, 
because they are close to each other (47.6 and 49.9 K; Table \ref{tb:lines}). 

{\bfresubfirst 
In determining the column densities and fractional abundances of SO and \SO\ relative to \hydro, 
we assume local thermodynamic equilibrium (LTE), 
as shown in Table \ref{tb:abundance}. 
We determine the column density and fractional abundance of \SO\ by assuming optically thin emission, 
and determine those of SO from the \SO\ results 
by using 
{\bfresubsecond $^{32}$S/$^{34}$S} 
ratio of 22.6. 
}

{\bfresubfirst 
The \SOO\ emission is intense around the protostar, 
while the SiO emission is marginally detected with a signal-to-noise (S/N) ratio of $\sim 5$ 
(Figure \ref{fig:dist_blowup}). 
The column densities and fractional abundances of \SOO\ and SiO 
are calculated under the same assumption as in the \SO\ case (Table \ref{tb:abundance}). 
}

The SO emission is also seen {\bfresubfirst on} the southeastern side of the continuum peak with an angular offset of about 4\arcsec\ ($\sim$500 au). 
This component has {\bfresubfirst an} intensity peak position of: 
$(\alpha_{2000}, \delta_{2000}) = (16^{\rm h}27^{\rm m}09\fs489\pm0.005, -24\degr37^\prime23\farcs48\pm0.05)$.
Its image component size is 
$\sim 2\farcs2 \times 1\farcs5\ (\sim 300\ {\rm au} \times 200\ {\rm au})$. 
{\bfresubfirst In addition,} the \SO\ and {\bfresubfirst \SOO\ emissions are} marginally seen around this position. 
This component could be due to 
{\bfresubfirst a local density enhancement} (see Section \ref{sec:dist_mol_CS}), 
as previously reported \citep{Lommen2008}. 
The {\bfresubfirst SO and \SOO\ column} densities are {\bfresubfirst obtained} from their emissions by assuming LTE 
and optically thin condition {\bfresubfirst (Table \ref{tb:comparison}).}  
We {\bfresubfirst assume a} gas temperature of 20 K {\bfresubfirst at} this position, 
according to the dust temperature reported by \citet{RochaPilling2018}; 
the dust temperature is about 25 K at {\bfresubfirst a} distance of 500 au from the protostar. 

\subsubsection{CS} \label{sec:dist_mol_CS} 
{\bfresubfirst In contrast} to the molecular emission described in the previous subsection, 
the CS emission is weak and marginally detected at the continuum peak position (Figure \ref{fig:dist_blowup}(b)). 
The faint emission of CS is also confirmed {\bfresubfirst by} its {\bfresubfirst spectrum} (Figure \ref{fig:spectra}), 
{\bfresubfirst which} shows a weak absorption feature with a narrow line width 
at {\bfresubfirst a} velocity of $\sim$5.5 \kmps\, probably due to foreground gas. 
{\bfresubsecond The results for} column density and fractional abundance of CS toward the continuum peak are {\bfresubsecond listed} in Table \ref{tb:abundance}. 
They 
might be {\bfresubfirst underestimates owing} to the marginal absorption in the red-shifted component. 
Nevertheless, CS is obviously deficient in the gas 
{\bfresubfirst near the protostar} in comparison {\bfresubfirst with} SO.

On the other hand, 
the CS emission is rather intense in the 
{\bfresubfirst ridge} component {\bfresubfirst on} the southeastern side (Figure \ref{fig:dist}(b)). 
This component traced by CS {\bfresubfirst extends} over 5\arcsec\ ($\sim$600 au) 
along the northeast-southwest direction (Figure \ref{fig:dist}(b)). 
The {\bfresubfirst CS} column density {\bfresubfirst of} this component is {\bfresubfirst determined} 
in the same way as {\bfresubfirst for} SO and \SOO\ (see Section \ref{sec:dist_mol_SO}). 
The results are shown in Table \ref{tb:comparison}. 

\subsection{Abundance Ratios of the S-bearing Species} \label{sec:dist_ratio}

{\bfresubfirst Table \ref{tb:comparison}} 
{\bfresubsecond lists the results for} CS, SO, and \SOO\ {\bfresubsecond abundance ratios.} 
{\bfresubfirst We have quantitatively confirmed that CS is much less abundant 
than SO and \SOO\ at the continuum peak position.}
We hereafter assume {\bfresubfirst a} gas temperature of 100 K for 
the continuum peak, 
{\bfresubfirst {\bfresubsecond based on} the dust temperature {\bfresubfirst distribution} reported by \citet{RochaPilling2018}.} 

Table \ref{tb:comparison} also shows 
{\bfresubfirst previously reported results} 
for comparison: 
the {\bfresubsecond shocked} {\bfresubfirst outflow} 
in L1157 \citep[L1157 B1;][]{Bachiller_L1157}, 
the protostar and the outflow of NGC1333 IRAS 2 \citep{Wakelam2005}, 
and the envelope gas of {\bfresubfirst the} low-mass protostellar source \iras\ Source B \citep{Drozdovskaya_PILS}. 
In low-mass protostellar sources, 
SO is generally thought to be abundant in shocked {\bfresubfirst regions.} 
Indeed, the \molratio{SO}{CS} ratio in L1157 {\bfresubfirst B1} is higher than that in \iras\ Source B by more than one order of magnitude. 
The results in NGC 1333 IRAS 2 show a large variation in this abundance ratio, 
which tends to be high in the shocked gas of the eastern outflow lobe \citep{Wakelam2005}. 
{\bfresubfirst Nevertheless,} 
Elias 29 obviously shows a {\bfresubfirst higher} \molratio{SO}{CS} {\bfresubfirst ratio}
{\bfresubfirst than these sources,} 
although the error is large; 
the ratio in Elias 29 is higher than that in L1157 B1 by a factor of $\sim$200 
and {\bfresubfirst higher} than the highest ratio in the shocked gas in NGC 1333 IRAS 2 by a factor of 7. 

{\bfresubfirst As shown in} Table \ref{tb:comparison}, 
\molratio{\SOO}{CS} 
{\bfresubfirst is also higher} in the shocked region {\bfresubfirst of} L1157 B1 {\bfresubfirst than in \iras\ Source B.} 
{\bfresubfirst As in the case of \molratio{SO}{CS}, 
the \molratio{\SOO}{CS} ratio} 
in Elias 29 is higher than in {\bfresubfirst the} other sources by two {\bfresubfirst orders} of magnitude. 

These results suggest that 
Elias 29 is significantly {\bfresubfirst richer} in SO and {\bfresubfirst \SOO\ than in CS.} 
Such a peculiar chemical characteristic is also reported for 
the low-mass Class I source LFAM 1 (or GSS 30 IRS 3) 
by \citet{Reboussin2015}; 
{\bfresubfirst that source} shows strong {\bfresubfirst emissions} of SO, \SOO, and SO$^+$. 
LFAM 1 is also located in the $\rho$ Ophiuchi star-forming region 
and has strong radio emission at 6 cm \citep{Leous1991}. 
\citet{Reboussin2015} interpreted their observations 
{\bfresubfirst as shock chemistry caused by} 
an outflow from {\bfresubfirst this} Class I source or from a {\bfresubfirst nearby young stellar} object. 
{\bfresubfirst However,} the SO and {\bfresubfirst \SOO\ emissions} in Elias 29 {\bfresubfirst are} clearly associated with the protostar, 
although {\bfresubfirst we cannot rule out a} contribution {\bfresubfirst from} the outflow shock near the launching point. 

The column densities {\bfresubfirst listed in Table \ref{tb:abundance}} 
are beam-averaged. 
As mentioned in Section \ref{sec:dist_mol}, 
the compact emission concentrated at the continuum peak is only marginally resolved. 
Thus, the {\bfresubfirst calculated} column densities {\bfresubfirst may} be {\bfresubfirst underestimates} 
and could be affected by different beam {\bfresubfirst dilutions} among the molecules. 
If the CS emission {\bfresubfirst is more diluted} than {\bfresubfirst that of} SO {\bfresubfirst or} \SOO, 
the \molratio{SO}{CS} and \molratio{\SOO}{CS} ratios could be overestimated at the CS peak. 
{\bfresubsecond Observations} at higher resolution {\bfresubsecond are required} 
to study such small-scale variation {\bfresubfirst ({\bfresubsecond on size scales of} a few 10s of au).}

Table \ref{tb:comparison} also shows the molecular abundances of CS, SO, and \SOO\ 
and {\bfresubfirst their relative} abundance ratios 
in the ridge component of Elias 29. 
In the ridge, 
CS seems to be more abundant than at the continuum peak, 
as shown in Figures \ref{fig:dist} and \ref{fig:dist_blowup}. 
The SO emission is clearly detected in the ridge as well, 
while the \SOO\ emission is {\bfresubsecond marginally} detected with {\bfresubfirst a} peak intensity of {\bfresubfirst about the} 3$\sigma$ {\bfresubfirst confidence level.} 
The \molratio{SO}{CS} and \molratio{\SOO}{CS} ratios 
are {\bfresubfirst lower} than those at the continuum peak by a factor of 5 to 8, 
although they are still high {\bfresubfirst compared} with {\bfresubfirst those} of the low-mass protostellar source 
\iras\ Source B \citep{Drozdovskaya_PILS}.



\section{Chemical Inventory} 
\label{sec:discuss_chem}

As discussed above, 
Elias 29 is rich in SO and \SOO, 
while deficient in CS. 
{\bfresubfirst Furthermore,} Elias 29 is deficient in saturated organic molecules. 
Figures \ref{fig:mom0_Cbearing} and \ref{fig:spectra_Cbearing} show 
the integrated intensity maps and spectra of 
the \MF\ (\mf), \DME\ (\dme), CCH (\cch), and \cycCH\ (\cycch) lines 
{\bfresubfirst toward the protostar.} 
These molecular lines are not detected in {\bfresubfirst our} observation.

For comparison, 
Figure \ref{fig:spectra_Cbearing} shows the spectra observed toward NGC1333 IRAS 4A2 with ALMA \citep{LopezSepulcre2017}. 
IRAS 4A is a binary system consisting of IRAS 4A1 and IRAS 4A2, 
where IRAS 4A2 is a hot corino 
and is chemically {\bfresubfirst richer than} IRAS 4A1. 
In IRAS 4A2, 
the organic molecular {\bfresubfirst lines } \MF, \DME, CCH, and \cycCH\ are {\bfresubfirst all} clearly detected, 
in contrast {\bfresubfirst with} the Elias 29 case, 
although the observational sensitivity is almost the same for the two sources. 

\subsection{Deficiency in Organic Molecules} \label{sec:discuss_chem_Cbearing}

\begin{figure}
	\iffigure
	\includegraphics[bb = 0 0 750 550, scale = 0.6]{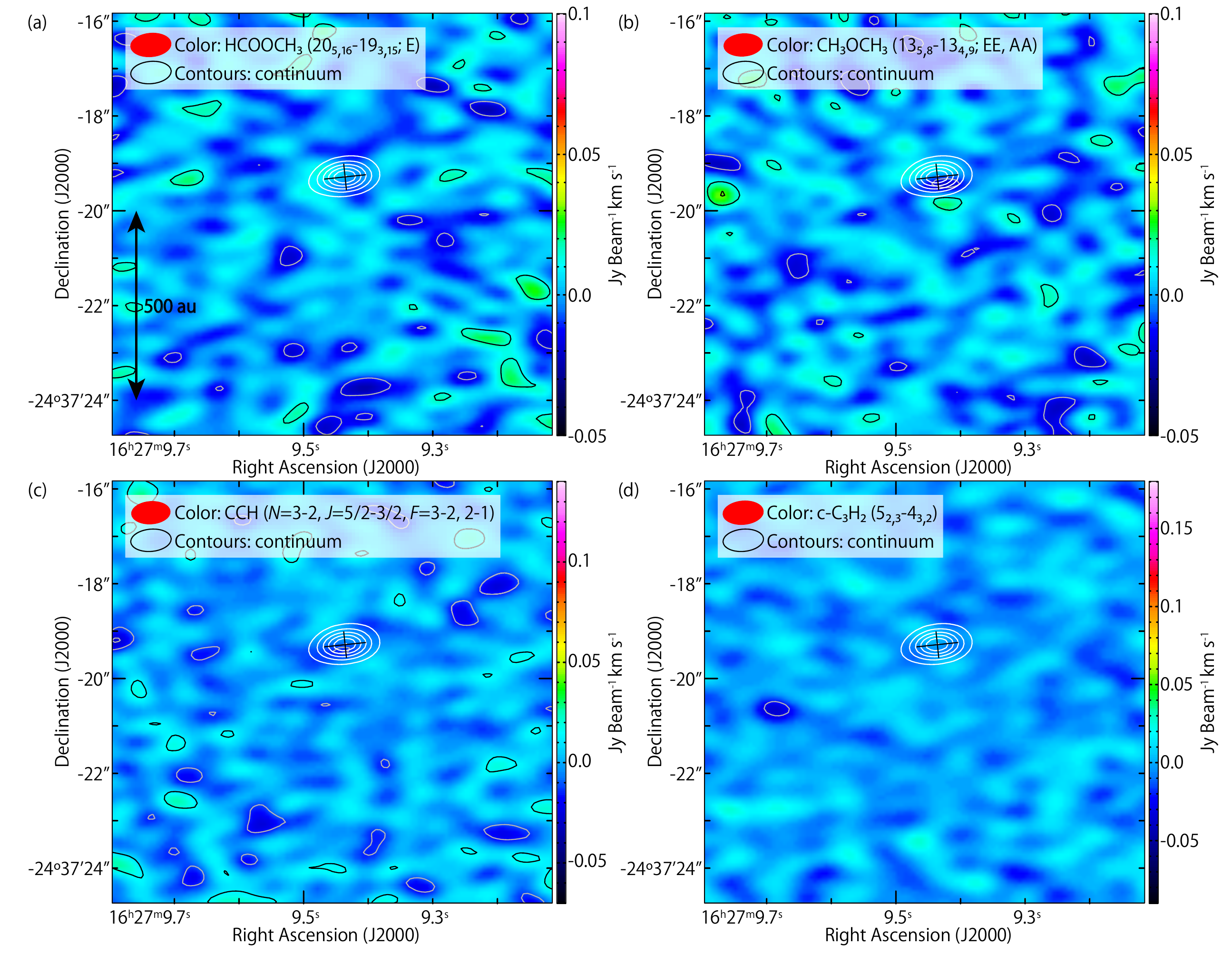}
	\fi
	\caption{Integrated intensity maps (color) of 
			the (a) \MF\ (\mf), (b) \DME\ (\dme), (c) CCH (\cch), and (d) \cycCH\ (\cycch) lines. 
			The velocity range for the integration is 1 to 7 \kmps. 
			{\bfresubfirst The} black cross in each panel represents 
			the continuum peak position and the FWHM widths 
			obtained {\bfresubfirst from} two-dimensional Gaussian fitting. 
			{\bfresubfirst The white} contours around the cross represent the 1.2 mm dust continuum emission map, 
			where the contour levels are the same as those in Figure \ref{fig:dist}. 
			{\bfresubfirst The black} and grey contours in each panel represent 
			the intensity map of each molecular line. 
			The contour levels are {\bfresubfirst at intervals of 3$\sigma$ starting from 3$\sigma$,} 
			where the rms {\bfresubfirst levels are} 
			5, 5, 7, and 9 m\Jypb\ \kmps\ for the \MF, \DME, CCH, and {\bfresubfirst \cycCH\ lines,} respectively. 
			\label{fig:mom0_Cbearing}}
\end{figure}

We here derive the upper limits to the column densities of the above molecules. 
The upper limits to the column densities 
{\bfresubfirst of \MF, \DME, CCH, and \cycCH,} 
are 
{\bfresubfirst determined from} 
the rms noise levels of {\bfresubfirst their} integrated intensity maps in Figure \ref{fig:mom0_Cbearing}, 
{\bfresubfirst as summarized in Table \ref{tb:abundance_Cbearing}.} 
{\bfresubfirst These upper limits are compared with the corresponding fractional abundances or their upper limits 
reported for NGC1333 IRAS 4A1 and A2 by \citet{LopezSepulcre2017} (Table \ref{tb:abundance_Cbearing}).} 
the upper limits {\bfresubfirst to} the abundances of these molecules in Elias 29 are  
lower than those in NGC1333 IRAS 4A2 by one order of magnitude, 
{\bfresubfirst while} the constraints for the fractional abundances in Elias 29 are 
{\bfresubfirst looser} 
than those in NGC 1333 IRAS 4A1.

{\bfresubfirst Because the emissions} {\bfresubsecond from} saturated organic molecules are expected 
to originate {\bfresubfirst from} the hot ($> 100$ K) region near the protostar, 
{\bfresubfirst they} could be weak if the hot region is small. 
This possibility is suggested for NGC1333 IRAS 4A1 \citep{LopezSepulcre2017}. 
{\bfresubfirst However, this} is not the case for Elias 29; 
according to \citet{Boogert2000}, 
the \Water\ gas, 
which is expected to reside in the hot ($> 100$ K) region, 
{\bfresubsecond was} detected in absorption {\bfresubsecond in this source by ISO,} 
with {\bfresubfirst a} column density of $\sim 10^{18}$ \cmsquare. 
{\bfresubfirst This} corresponds to a high fractional ratio of $10^{-5}$ {\bfresubsecond relative} to the total hydrogen column density 
($N_{\rm H} \sim 10^{23}$ \cmsquare). 
{\bfresubfirst According to \citet{Boogert2000},} 
Elias 29 should have a hot region 
where various saturated organic molecules {\bfresubfirst are} liberated. 
They also reported the size of the hot core to be $(85-225)$ au 
{\bfresubsecond based on observations of} 
infrared absorption of high-$J$ vibration-rotation lines of CO. 
{\bfresubfirst This size can be resolved in our observation {\bfresubfirst with our} beam size $\sim$ 100 au).} 
Furthermore, the dust temperature is above 100 K within $\sim 50$ au {\bfresubfirst of} the protostar 
according to \citet{RochaPilling2018}. 
{\bfresubsecond Therefore, the non-detection of COM lines in our observations 
is not due to their {\bf being} frozen-out; 
{\bf instead,} it is likely that Elias 29 is {\bf very} poor in COMs.}

The integrated intensity map of the CCH line (Figure \ref{fig:mom0_Cbearing}c) 
shows a slight negative intensity with a minimum value of 
$-91$ m\Jypb\ \kmps\ ($\sim$6$\sigma$) {\bfresubfirst around} the protostar, 
This feature is likely due to self-absorption by foreground gas containing CCH. 
The extended ambient gas would be resolved-out in this observation with the interferometer, 
whose maximum recoverable size is $\sim$6\arcsec. 
Nevertheless, 
no enhancement of {\bfresubfirst the} \MF, \DME, CCH, 
or {\bfresubsecond \cycCH\ emission}
{\bfresubfirst is} confirmed at least near the protostar, in contrast to the SO and \SOO\ cases.

\subsection{Possible Cause for the Chemical Characteristics of Elias 29} \label{sec:discuss_chem_highT}

As described above, 
one of the interesting chemical features of Elias 29 is its deficiency of organic molecules 
as well as its richness in 
{\bfresubfirst SO and \SOO.} 
Here we discuss the following two possibilities: 
an evolutionary effect after the protostellar birth and an environmental effect. 
{\bfresubsecond Distinguishing between these alternatives} is left 
for future {\bfresubfirst study.}

Elias 29 is a Class I protostar with {\bfresubfirst a} bolometric temperature of {\bfresubfirst 391 K \citep{Miotello2014}}. 
Hence, the disk component would have already experienced 
{\bfresubsecond a temperature above that corresponding to the sublimation of COMS, given} 
the age of {\bfresubfirst this} protostar 
\citep[$(1-5) \times 10^5$ yr after the onset of gravitational collapse;][]{Chen_AgefromTbol}. 
After sublimation, COMs are {\bfresubsecond destroyed} by proton transfer reactions with \HCO, H$_3$O$^+$, 
and other ions followed by electron recombination reactions. 
According to the hot-core model {\bfresubfirst proposed} by \citet{NomuraMillar2004}, 
the time scale for the 
{\bfresubsecond destruction} is about $10^5$ yr, 
which {\bfresubfirst is} comparable to the age of the Elias 29 protostar. 
In this case, 
the organic molecules liberated from dust grains at the birth of the protostar 
would have already been broken up by gas reactions. 
If most of the infalling envelope has already dissipated, 
{\bfresubfirst fresh grains with COM-ice are no longer supplied.} 
In this case, COMs {\bfresubfirst would} be deficient near the protostar. 
{\bfresubfirst As for the sulfur-bearing molecules, 
SO and \SOO\ can be abundant while CS can be deficient 
in the gas of an evolved source. 
SO and \SOO\ are} 
the most abundant sulfur-bearing molecules in the gas 
{\bfresubfirst when {\bfresubsecond the chemical composition of the gas} is in a} steady state. 
Thus, {\bfresubfirst the} relative {\bfresubfirst maturity} of Elias 29 can explain {\bfresubfirst its chemical characteristics.} 
To {\bfresubfirst test} this {\bfresubfirst hypothesis,} 
{\bfresubfirst it is necessary to observe} other evolved protostellar sources {\bfresubfirst at high spatial resolution and} 
to confirm the dissipation of the envelope gas, 
which is resolved out in the present observation.

Alternatively, 
both the deficiency of organic molecules and the richness in SO and \SOO\ 
in Elias 29 {\bfresubfirst could} be attributed to the relatively high temperature of the parent core in the starless-core phase 
due to external heating by nearby {\bfresubfirst objects. 
These objects include} YSOs 
{\bfresubfirst that are} only {\bfresubfirst about 3\arcmin\ from Elias 29,} 
as well as two bright B stars (S1 and HD147889) 
{\bfresubfirst \citep{Yui1993, Ebisawa2015, RochaPilling2018}.} 
As mentioned in Section \ref{sec:intro}, 
a warm parent core is consistent 
with the infrared observations of CO and \COO\ \citep{Boogert2000}. 
It is generally thought that saturated organic molecules, 
such as \MN\ and \MF, 
are produced on dust grains 
by hydrogenation of CO 
and liberated into the gas near the protostar. 
It {\bfresubfirst has} also {\bfresubfirst been} proposed that 
saturated organic molecules {\bfresubfirst are} produced in the gas from \MN\ 
liberated from dust grains \citep{Balucani2015_COMsinGas}. 
Meanwhile, it {\bfresubfirst has been} suggested that unsaturated carbon-chain molecules and related species, 
such as CCH and \cycCH, 
are efficiently produced from {\bfresubfirst methane (\MT)} {\bfresubsecond via} sa gas-phase reaction \citep{Sakai_ChemRev}. 
{\bfresubfirst \MT\ is} the precursor of unsaturated carbon chains 
{\bfresubfirst and} is mainly formed by hydrogenation of atomic carbon on dust grains. 
Therefore, 
depletion of either CO or atomic carbon {\bfresubfirst onto dust grains} in the prestellar-core stage is required 
for {\bfresubfirst enhanced production} of saturated and unsaturated organic molecules in the protostellar core. 
Both CO and atomic carbon are depleted onto dust grains 
with {\bfresubsecond temperatures} lower than 20 K 
(see Appendix \ref{sec:appendix_Tdes}). 

As mentioned in Section \ref{sec:intro}, 
the dust temperature of the protostellar core Elias 29 
is mostly higher than 20 K even {\bfresubfirst on the scale of a few thousand} au according to \citet{RochaPilling2018}. 
If the dust temperature {\bfresubfirst were} as high as 20 K 
in the parent cloud of Elias 29 in the past  
as {\bfresubfirst it is now,} 
CO and atomic carbon would hardly be adsorbed onto dust grains. 
This may result in insufficient production of the above organic molecules. 
If this is the case, 
the deuterium fractionation ratio, 
which increases after CO depletion onto dust grains, 
is expected to stay low 
{\bfresubfirst \citep[e.g.][]{Caselli2002, Bacmann2003, Crapsi2005},} 
although {\bfresubfirst evolutionary effects} should also be considered \citep{Imai2018}. 

Under the above temperature condition, 
sulfur atoms will not {\bfresubfirst be depleted} onto dust grains {\bfresubfirst either,} 
because the desorption temperature of sulfur atoms 
is comparable to {\bfresubfirst those} of CO and atomic carbon 
\citep[KIDA;][Appendix \ref{sec:appendix_Tdes}]{Wakelam_KIDA}. 
{\bfresubsecond In this case,} 
sulfur atoms {\bfresubfirst are} converted to SO and \SOO\ through gas-phase reactions 
{\bfresubfirst \citep[e.g.][]{Prasad1982, Charnley1997, Yoneda_CheminDisk, Wakelam2011}.} 
{\bfresubfirst High} \molratio{SO}{CS} and \molratio{\SOO}{CS} {\bfresubfirst ratios are} expected 
if the 
{\bfresubfirst elemental C/O ratio is $> 1$, 
which is plausible considering the detection of water vapor.} 

{\bfresubfirst Similar} chemical characteristics {\bfresubfirst are reported for two other sources: 
the} massive star-forming region G5.89--0.39 and {\bfresubfirst the} nearby low-metallicity 
Large Magellanic Cloud (LMC). 
G5.89--0.39 {\bfresubfirst consists} of shock-heated gas ($150-1800$ K) 
{\bfresubsecond based on interpretation of} CO observations
{\bfresubfirst with the SMA} tracing the outflows \citep{SuYN2012}. 
{\bfresubfirst This} source 
{\bfresubfirst has} intense {\bfresubfirst SO and \SOO\ line emissions 
observed by} \citet{Thompson1999} with {\bfresubfirst the} JCMT and {\bfresubfirst by} \citet{Hunter2008} with {\bfresubfirst the} SMA. 
\citet{Hunter2008} also found relatively weak \MN\ line {\bfresubfirst emission} 
and 
{\bfresubsecond suggested, as a possible cause,} 
that CO is not {\bfresubfirst well} adsorbed onto dust grains. 
In the high-mass star-forming core ST11 in {\bfresubfirst the LMC}, 
{\bfresubfirst {\bfresubsecond an} ALMA observation by \citet{Shimonishi2016} shows bright SO and \SOO\ and weak \MN\ line emissions.} 
While the low metallicity in the LMC \citep[e.g.,][]{Dufour1982}
should partly contribute to the low \MN\ abundance, 
{\bfresubsecond \citet{Shimonishi2016}} 
suggest that the \MN\ production is suppressed 
{\bfresubfirst by} warm ice chemistry at the molecular cloud stage. 
{\bfresubfirst This is supported by a chemical model \citep{Acharyya2018}.} 
{\bfresubfirst We} note that CS is less abundant in ST11 than {\bfresubfirst in hot cores in the Milky Way,} 
which {\bfresubfirst is} similar to the Elias 29 case. 
Elias 29 {\bfresubfirst has low mass,} 
while the above sources are {\bfresubfirst massive.} 
However, 
their similar chemical characteristics could be interpreted 
{\bfresubsecond in terms of} 
the same scenario 
with a relatively high dust temperature {\bfresubfirst during the prestellar core phase} discussed above.

In this regard, 
one 
{\bfresubsecond might conclude} 
that warm core {\bfresubfirst temperatures are common} 
even for high-mass star-forming regions harboring hot {\bfresubfirst cores} rich in COMs. 
However, this does not {\bfresubfirst necessarily} contradict the above discussion. 
A key factor {\bfresubfirst in} insufficient {\bfresubfirst COM} production on dust surfaces 
is not the dust temperature at the current protostellar core phase 
but {\bfresubsecond the temperature} 
during the prestellar phase. 
Therefore, high-mass star-forming regions rich {\bfresubfirst in} COMs can occur 
if their dust temperature {\bfresubfirst in the prestellar core phase} was cold enough for CO depletion 
and then {\bfresubfirst rose} to the current {\bfresubfirst level} after the 
{\bfresubsecond protostellar} birth.

\section{Rotation Traced by SO} \label{sec:kinstr}

\begin{landscape}
\begin{figure}
	\iffigure
	\includegraphics[bb = 0 0 800 300, scale = 0.7]{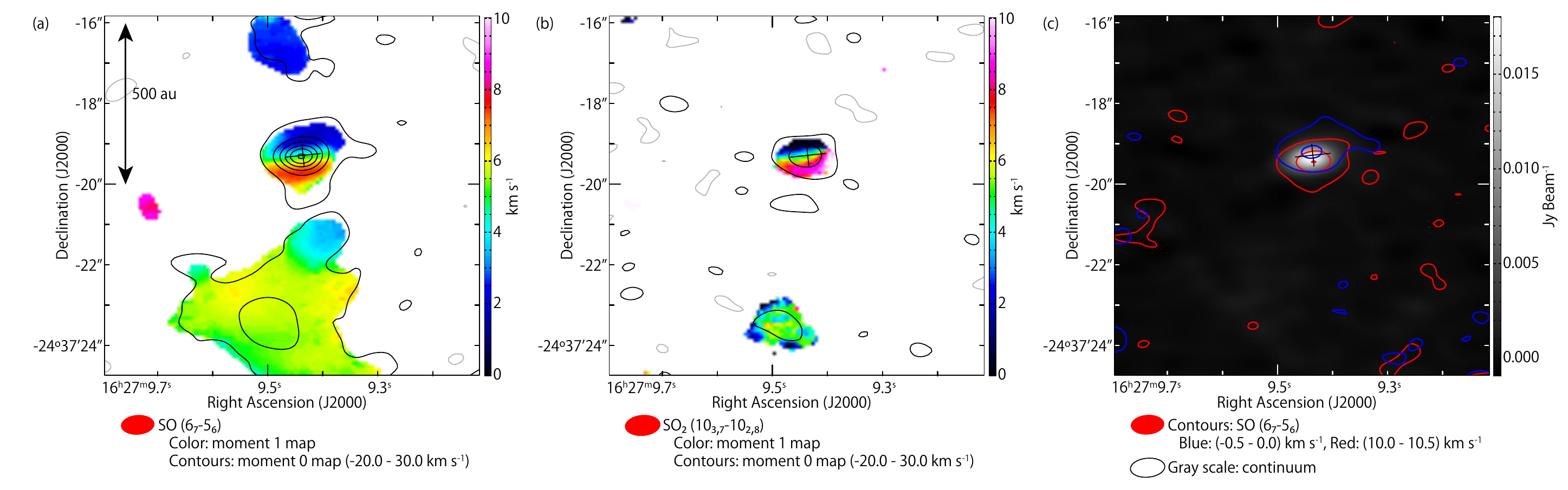}
	\vspace*{-25pt}
	\fi
	\caption{(a) Velocity map of the SO (\so) emission (color). 
			{\bfresubfirst The contours} represent the integrated intensity map of SO (Figure \ref{fig:dist_blowup}(c)). 
			{\bfresubfirst The contour} levels are {\bfresubfirst at intervals of 10$\sigma$ starting} from 2$\sigma$, 
			where the rms level is 60 m\Jypb. 
			(b) Velocity map of the \SOO\ (\soo) emission (color). 
			{\bfresubfirst The contours} represent the integrated intensity map of \SOO\ (Figure \ref{fig:dist_blowup}(e)). 
			{\bfresubfirst The contour} levels are {\bfresubfirst at intervals of 10$\sigma$ starting} from 2$\sigma$, 
			where the rms level is 30 m\Jypb. 
			(c) Integrated intensity maps of the SO emission (contours). 
			The gray scale represents the 1.2 mm continuum map. 
			The velocity ranges for integration are 
			$-0.5$ to 0.0 \kmps\ (blue contours) 
			and 10.0 to 10.5 \kmps\ (red contours). 
			{\bfresubfirst The contour} levels are {\bfresubfirst at intervals of 20$\sigma$ starting} from 2$\sigma$, 
			where the rms level is 3 m\Jypb. 
			{\bfresubfirst The} black cross in each panel represents the continuum peak position as in Figure \ref{fig:dist}. 
			\label{fig:SO-velmap}}
\end{figure}
\end{landscape}

Figure \ref{fig:SO-velmap}(a) shows the velocity map {\bfresubfirst (the moment 1 map)} of the SO (\so) line. 
It reveals a clear velocity gradient along the north-south direction across the continuum peak. 
This gradient is almost perpendicular to the {\bfresubfirst outflow} axis, 
which {\bfresubfirst was} previously reported 
{\bfresubsecond as running} 
along the east-west direction 
{\bfresubfirst near the protostar} \citep{Ceccarelli2002}. 
Thus, the gradient most likely represents {\bfresubsecond rotation.} 
Considering its compact distribution, 
{\bfresubsecond the SO emission} 
seems to trace the \desys\ in the vicinity of the protostar. 
A similar velocity gradient around the protostar is also seen in the \SOO\ (\soo) emission (Figure \ref{fig:SO-velmap}b). 
The SO line has a faint blue-shifted component 
at {\bfresubfirst a} distance of 3\arcsec\ ($\sim 400$ au) north {\bfresubfirst of} the protostar, 
which may come from part of the gas {\bfresubfirst rotating} around the protostar. 
{\bfresubfirst This velocity structure is consistent with 
the observation of HCO$^+$ ($J=3-2$) reported by \citet{Lommen2008}.} 
{\bfresubfirst Meanwhile, 
no} clear velocity shift is seen 
in the SO emission in the ridge component. 
This component has a large velocity range of $(3-8)$ \kmps\ (a velocity-shift range of $(-1-+4)$ \kmps\ {\bfresubsecond with respect to the systemic velocity of Elias 29),} 
{\bfresubfirst which} 
{\bfresubfirst cannot} be attributed to {\bfresubfirst rotation} around the protostar. 
Alternatively, 
{\bfresubfirst the ridge} 
{\bfresubfirst may come} from a gas component 
in the complex geometrical system {\bfresubfirst surrounding} Elias 29 \citep{Boogert2002_Elias29}.

Figure \ref{fig:SO-velmap}(c) shows the integrated intensity maps of the high velocity-shift components of the SO emission. 
The velocity ranges for the integration {\bfresubfirst are} $-0.5$ to 0.0 \kmps\ and 10.0 to 10.5 \kmps\ for {\bfresubfirst the} 
{\bfresubfirst blue- and red-shifted components, respectively.} 
{\bfresubfirst Two-dimensional Gaussian fitting yields the} intensity peak positions  
$(\alpha_{2000}, \delta_{2000}) = (16^{\rm h}27^{\rm m}09\fs4353\pm0.0011, -24\degr37^\prime19\farcs145\pm0.013)$ and 
$(\alpha_{2000}, \delta_{2000}) = (16^{\rm h}27^{\rm m}09\fs4343\pm0.0008, -24\degr37^\prime19\farcs419\pm0.004)$ 
for the blue- and red-shifted components, respectively. 
Although the separation ($\sim$0\farcs3; $\sim$30 au) is marginal, 
these {\bfresubfirst peaks} are on opposite sides {\bfresubfirst of} the continuum peak. 
Moreover, 
they align almost on a common line with the {\bfresubfirst P.A.} of $\sim$0\degr. 
{\bfresubfirst On the basis of} this result, 
we define the P.A. of the mid-plane of the \desys\ to be 0\degr, 
{\bfresubfirst which means} the P.A. of the rotation axis {\bfresubfirst is 270\degr}. 

The image component size of the integrated intensity map of the SO line deconvolved from the {\bf storing} beam 
is $\sim 0\farcs7 \times 0\farcs3$ ($\sim 100 \times 40$ au) (P.A. $158\degr \pm 11\degr$) 
(see Section \ref{sec:dist_mol_SO}). 
{\bfresubfirst If we assume} a flat disk with {\bfresubfirst no} thickness, 
$i$ is estimated to be 65\degr. 
When the thickness of the disk is considered, 
this value is regarded as the lower limit. 
This result is 
in contrast to the {\bfresubfirst \ia\ of} less than 60\degr\ reported by \citet{Boogert2002_Elias29} 
{\bfresubfirst on the basis of a flat spectral energy distribution (SED).
However, the analysis of the SED could be affected by 
{\bfresubsecond gas in the foreground} 
of Elias 29 \citep[e.g.][]{Boogert2002_Elias29, RochaPilling2018}, 
and thus our result does not seriously conflict with their estimate. 
For a further constraint, 
detailed analysis of the outflow will be helpful.}

\subsection{Kinematic Structure around the Protostar} \label{sec:kinstr_pv}

\begin{figure}
	\iffigure
	\includegraphics[bb = -50 0 500 680, scale = 0.75]{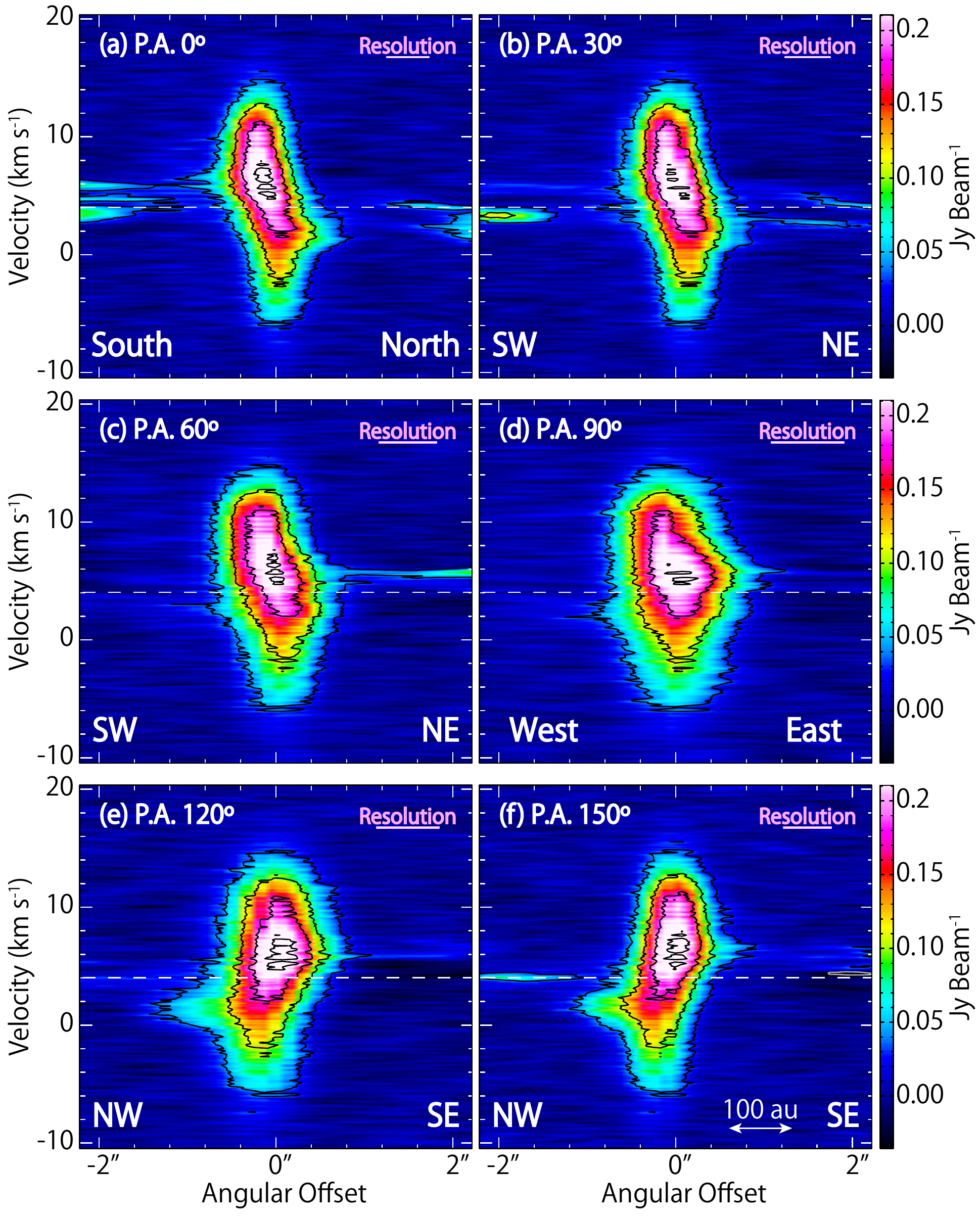}
	\vspace*{-20pt}
	\fi
	\caption{Position-velocity diagrams of the SO (\so) line. 
			The position axes are centered at the continuum peak 
			and are along the position {\bfresubfirst angles} 
			(a) 0\degr, (b) 30\degr, (c) 60\degr, (d) 90\degr, (e) 120\degr, and (f) 150\degr. 
			{\bfresubfirst The} position axis in panel (a) is along the mid-plane of the \desys. 
			{\bfresubfirst The white} dashed lines represent the \sysV\ of \vsysval\ \kmps. 
			\label{fig:PV_SO}}
\end{figure}

Figure \ref{fig:PV_SO} shows position-velocity (PV) diagrams of the SO line. 
The position axes are centered at the continuum peak position. 
{\bfresubsecond The P.A.s of the position axes} 
are taken for every 30\degr\ 
{\bfresubfirst and} are shown in Figure \ref{fig:dist_blowup}(c); 
Figure \ref{fig:PV_SO}(a) 
{\bfresubsecond shows the PV diagram} 
along the mid-plane of the \desys\ (P.A. 0\degr), 
while Figure \ref{fig:PV_SO}(d) is along the line perpendicular to it (P.A. 90\degr). 

Figure \ref{fig:PV_SO}(a) 
{\bfresubfirst shows} a bar-like feature with a clear velocity gradient across the continuum peak; 
the SO emission is blue- and red-shifted {\bfresubfirst on} the northern and southern sides of the continuum peak, respectively. 
This velocity gradient corresponds to that in Figure \ref{fig:SO-velmap}. 
The velocity gradient becomes less clear as {\bfresubfirst the} P.A. of the position axis {\bfresubfirst increases} from 0\degr\ to 90\degr, 
and it is hardly seen in Figure \ref{fig:PV_SO}(d) (P.A. 90\degr). 
In the PV diagrams with {\bfresubfirst a P.A. greater} than 90\degr\ (Figures \ref{fig:PV_SO}(e) and (f)), 
the velocity gradient is confirmed again; 
the {\bfresubfirst emissions on} the northwestern and southeastern sides of the continuum peak 
{\bfresubfirst are} blue- and red-shifted, respectively. 
These features are 
{\bfresubsecond most likely} attributed to {\bfresubfirst rotation} along the north-south direction 
without significant infall motion. 

\begin{figure}
	\iffigure
	\includegraphics[bb = -50 0 500 680, scale = 0.75]{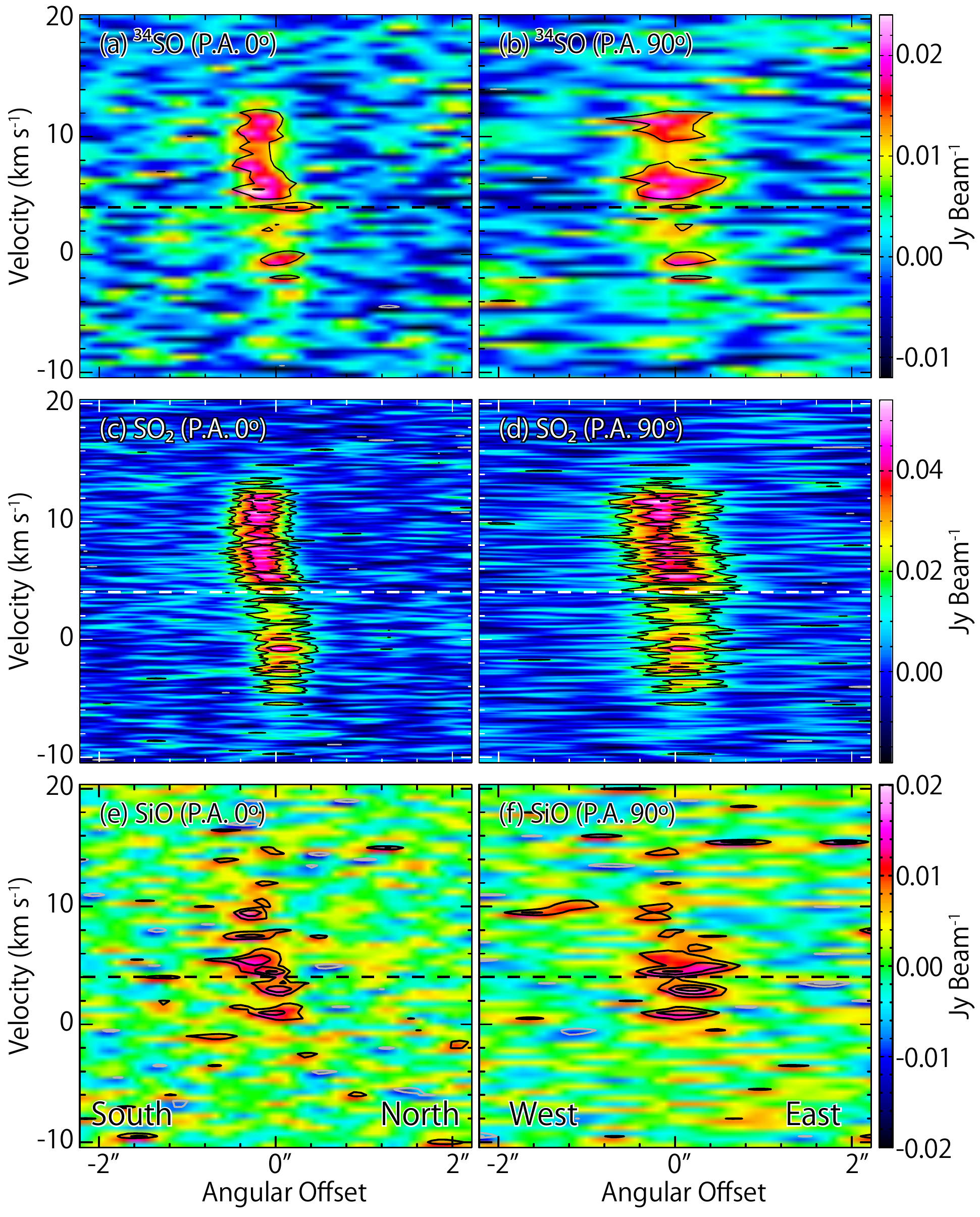}
	\vspace*{-20pt}
	\fi
	\caption{Position-velocity diagrams of the (a, b) \SO\ (\tfso), (c, d) \SOO\ (\soo), and (e, f) DeleteSiO (\sio) lines. 
			The position axes of panels (a, c, e) are the same as {\bfresubfirst those} in panel (a) {\bfresubfirst of} Figure \ref{fig:PV_SO}, 
			and those of panels (b, d, f) are the same as {\bfresubfirst those} in panel (d) {\bfresubfirst of} Figure \ref{fig:PV_SO}. 
			{\bfresubfirst The black} and white dashed lines represent {\bfresubfirst a} \sysV\ of \vsysval\ \kmps. 
			\label{fig:PV_34SO-SO2-SiO}} 
\end{figure}

Figure \ref{fig:PV_34SO-SO2-SiO} 
shows the PV diagrams of the \SO, \SOO, and SiO lines. 
The S/N ratio is worse for these lines than for the SO line. 
Nevertheless, 
the velocity gradient along the north-south direction is {\bfresubfirst as} clear in the \SOO\ emission (Figure \ref{fig:PV_34SO-SO2-SiO}c), 
as {\bfresubfirst in} the SO emission. 
It is difficult to confirm the velocity gradient in the \SO\ and SiO emissions 
because of their insufficient S/N ratios, 
although 
{\bfresubsecond a similar velocity-gradient} 
{\bfresubfirst is marginally} seen in the SiO emission. 
Moreover, the red-shifted components look more intense than the blue-shifted components 
in the SO and {\bfresubfirst \SOO\ emissions,} and maybe also in the \SO\ emission. 
This is discussed {\bfresubfirst below} in Section \ref{sec:discuss_spec}.

\subsection{Analysis of Rotation} \label{sec:kinstr_Kep}

{\bfresubfirst In this section, 
we discuss the rotation around the protostar found in the SO and \SOO\ lines. 
Because the rotational structure is marginally resolved as shown in Figures \ref{fig:SO-velmap} and \ref{fig:PV_SO}, 
it is difficult to distinguish between Keplerian motion and infall-rotation. 
Although a clear infall motion is not apparently seen in the PV diagram (Figure \ref{fig:PV_SO}), 
we cannot exclude the {\bfresubsecond existence of} infall at the current stage. 
Moreover, the SO and \SOO\ spectra do not show the symmetric double-peak characteristic of Keplerian motion 
\citep[e.g.][]{Eracleous1994, Dutrey1997, Oberg2015, Kastner2018, Imai2019}, 
which suggests a contribution from infall. 
Observations at higher angular resolution are thus necessary to determine the kinematics 
of the disk/envelope structure, including the disk size and the protostellar mass, accurately. 
Nonetheless, it is worthwhile to estimate the protostellar mass 
on the basis of the rotational structure observed in our study. 
For this purpose, we hereafter assume that the observed rotation is Keplerian, for simplicity.}

\begin{figure}
	\iffigure
	\includegraphics[bb = -50 0 500 680, scale = 0.75]{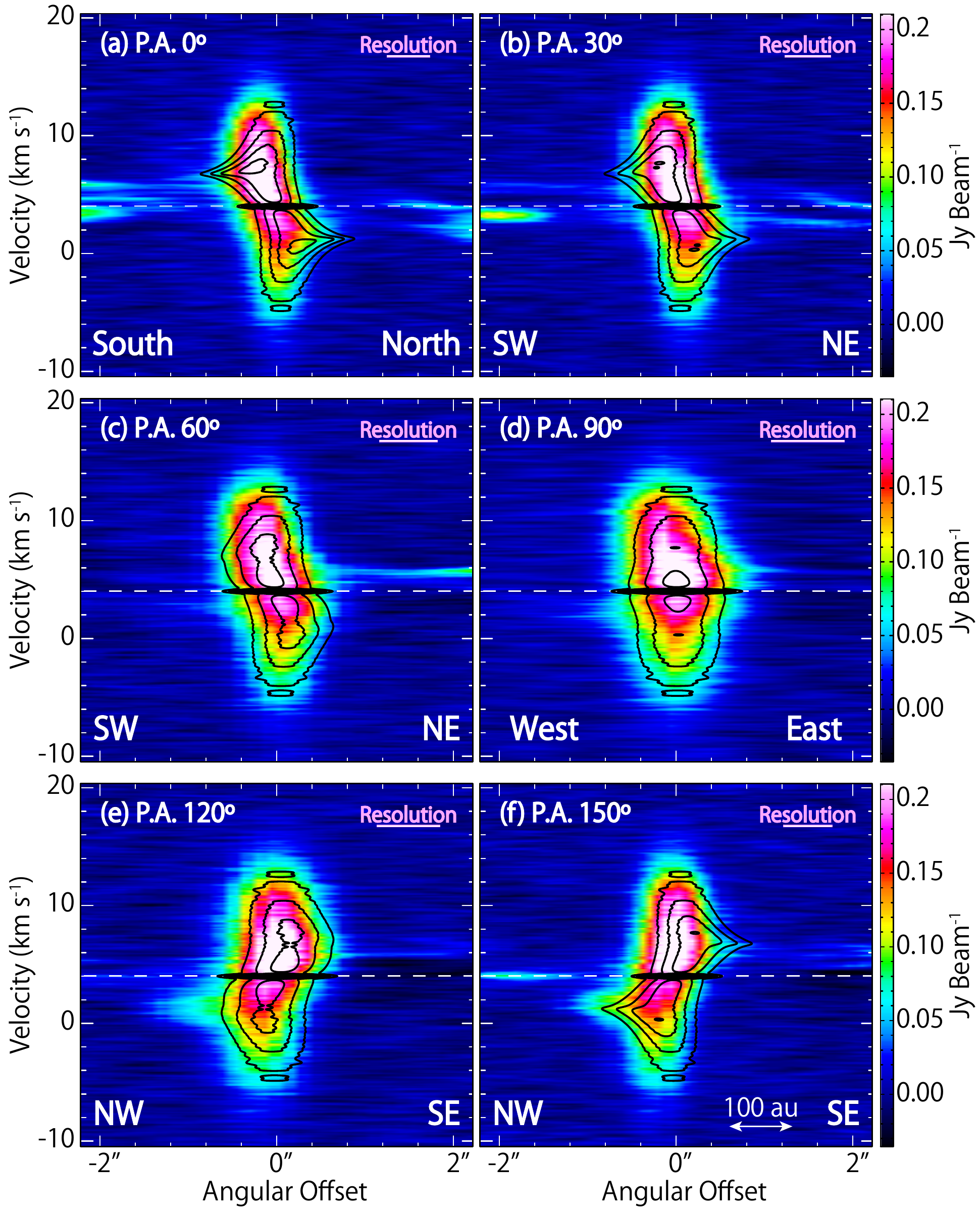}
	\vspace*{-20pt}
	\fi
	\caption{Position-velocity diagrams of the SO (\so; color) {\bfresubfirst line} in Figure \ref{fig:PV_SO} are 
			compared with the result of the Keplerian disk model (black contours). 
			{\bfresubfirst The position axes are the same as those in Figure \ref{fig:PV_SO}. 
			{\bfresubfirst The white} dashed lines represent {\bfresubfirst a} \sysV\ of \vsysval\ \kmps. 
			We assume}
			a protostellar mass of 1.0 \Msun\ and {\bfresubfirst \ia\ of 
			{\bf 65\degr\ (\incRem)} for the \desys. 
			In addition,} 
			the emission is assumed to come from the compact region 
			around the protostar with a radius of {\bfresubsecond 100 au.} 
			{\bfresubfirst The contour} levels for the Keplerian disk model 
			{\bfresubsecond are at intervals of 5\%\ starting from 5\%\ of the peak intensity.} 
			\label{fig:PV_SO-Kep}}
\end{figure}

Figure \ref{fig:PV_SO-Kep} shows the PV diagrams 
{\bfresubfirst simulated by using} the Keplerian disk model (contours); 
{\bfresubfirst the simulated diagrams} are superposed on those of the observed SO emission (color). 
We {\bfresubfirst use a proportionality} coefficient of 
{\bfresubsecond $r^{-2.0}$ for the emissivity}, 
including the {\bfresubfirst effects} of the molecular abundance and temperature profiles, 
where $r$ denotes the distance from the protostar. 
In this model, 
we ignore radiative transfer for simplicity. 
Thus, 
the {\bfresubfirst simulated} intensity distribution is not accurate enough 
for detailed comparison with the observation {\bfresubfirst owing} to systematic errors. 
Nevertheless, this simplified model is useful, as {\bfresubfirst found} for other sources 
\citep[e.g.][]{Oya_483, Okoda2018}, 
if we focus on the velocity profiles in the comparison. 
The emission in the model is convolved with the {\bf storing} beam in the {\bfresubfirst observed} SO line.

In Figure \ref{fig:PV_SO-Kep}, 
the following parameters are {\bfresubfirst used} for the Keplerian disk model: 
the protostellar mass is 1.0 \Msun, the \ia\ of the \desys\ is {\bf 65\degr\ (\incRem),} 
and the mid-plane of the disk is extended along {\bfresubfirst a} P.A. of 0\degr. 
The emission is assumed to come from the compact region 
around the protostar with a radius of {\bfresubsecond 100 au.} 
This model seems to 
{\bfresubsecond roughly} explain the observed velocity structure, 
{\bfresubfirst namely,} the velocity gradient in the PV diagrams of the SO line. 
When {\bfresubsecond the} \ia\ $i$ is {\bfresubfirst considered} explicitly, 
the protostellar mass {\bfresubfirst is given by} 
\begin{equation}
	M = {\bfresubsecond 0.82}\ M_\odot \times \dfrac{1}{\sin^2 i}, \label{eq:mass} 
\end{equation}
where {\bfresubfirst $i =$ 0 for a face-on configuration.} 
For instance, 
{\bfresubsecond the upper limit of $M$ could be 1.0 \Msun\ for the lower limit of $i$ (65\degr; Section \ref{sec:kinstr}), 
while the lower limit could be 0.82 \Msun\ for the completely edge-on case ($i = 90$\degr).} 
{\bfresubsecond Although these values are not contradict with the lower limit of 0.62 \Msun\ reported by \citet{Lommen2008} 
on the basis of} 
their 1.1 mm SMA data 
{\bfresubfirst of the \HCO\ ($J = 3-2$) line at a resolution of $4\farcs0 \times 2\farcs3$,} 
{\bfresubsecond assuming the edge-on configuration, 
the values} 
{\bfresubfirst are lower than} 
{\bfresubsecond those previously employed by a factor of a few;} 
for example, 
\citet{Lommen2008} reported {\bfresubfirst $2.5 \pm 0.6$ \Msun\ for an \ia\ of 30\degr,} 
whereas \citet{Miotello2014} reported {\bfresubfirst 3 \Msun\ using 
a continuum model. 
Our result is based on the kinematic structure observed at much higher angular resolution than in the previous studies 
and provides a better estimate. 
{\bfresubsecond If the protostellar mass is as small as $\leq 1.0$ \Msun, 
the protostellar age \citep[a few $10^5$ yr;][]{Lommen2008} 
would be estimated to be younger by a factor of a few.} 
The discrepancy between previous studies and ours is mainly due to the different \ia\ assumed.
Indeed, the protostellar mass could be {\bfresubsecond calculated to be 3.3} 
\Msun\ {\bfresubsecond with the equation (\ref{eq:mass}),} 
if the \ia\ is 30\degr\ as assumed by \citet{Lommen2008}. 
A tighter constraint on the protostellar mass {\bfresubsecond and the \ia} 
would require higher angular resolution.} 

{\bfresubfirst We note that 
our estimate of the protostellar mass can vary by a factor of a few 
when we consider possible infall motion.}
For instance, if we {\bfresubfirst used} the \ire, 
the protostellar mass would be half of 
{\bfresubsecond that estimated above via the pure} Keplerian model
(See Appendix \ref{sec:appendix_mass}). 

\subsection{Asymmetric Spectral Line Profiles of the \SOO\ emission} \label{sec:discuss_spec}

\begin{figure}
	\iffigure
	\includegraphics[bb = -250 0 600 800, scale = 0.65]{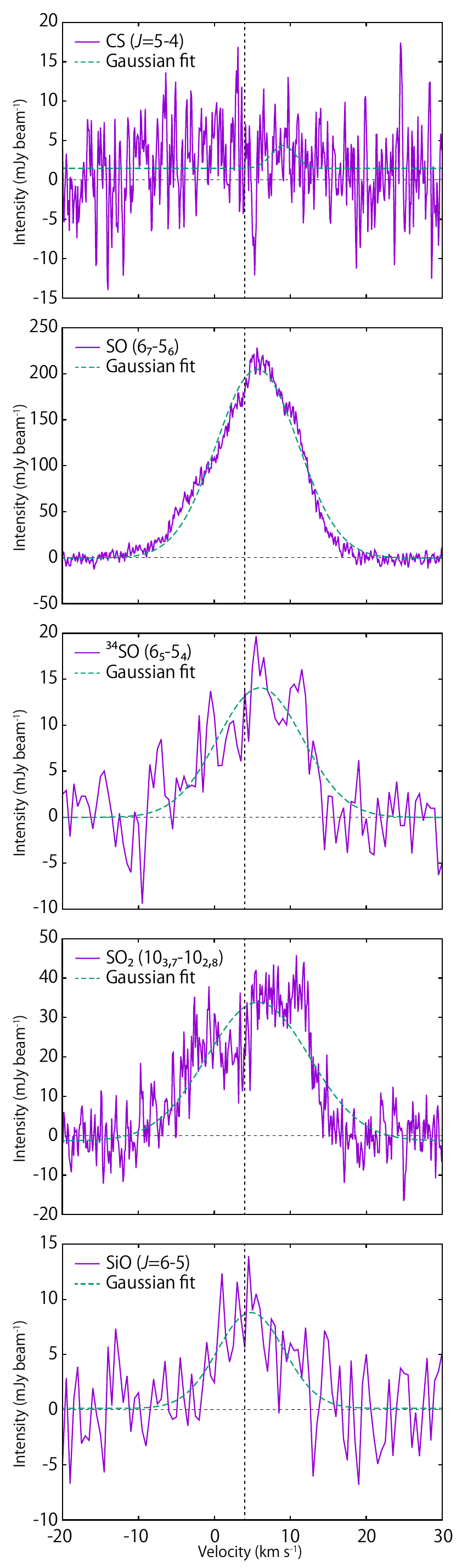}
	\vspace*{-30pt}
	\fi
	\caption{Spectra of the \alllines\ lines 
			averaged {\bfresubfirst over} a circular region centered at the continuum peak with a diameter of 0\farcs5. 
			{\bfresubfirst The} green dashed line in each panel shows the result of the Gaussian fitting. 
			{\bfresubfirst The} line widths (FWHM) obtained {\bfresubfirst from the Gaussian fitting are} 
			$12.7 \pm 0.1$, $13.3 \pm 1.3$, $15.9 \pm 0.6$, and $10.4 \pm 1.6$ \kmps\ 
			for the \linesexceptCS\ lines, respectively, 
			{\bfresubfirst where} 
			the central velocities are {\bfresubfirst calculated to be} 
			$5.65 \pm 0.04$, $6.02 \pm 0.45$, $5.76 \pm 0.18$, and $4.80 \pm 0.61$ \kmps, {\bfresubfirst respectively.}
			\label{fig:spectra}}
%
%
\end{figure}

\begin{figure}
	\iffigure
	\includegraphics[bb = -80 0 600 800, scale = 0.65]{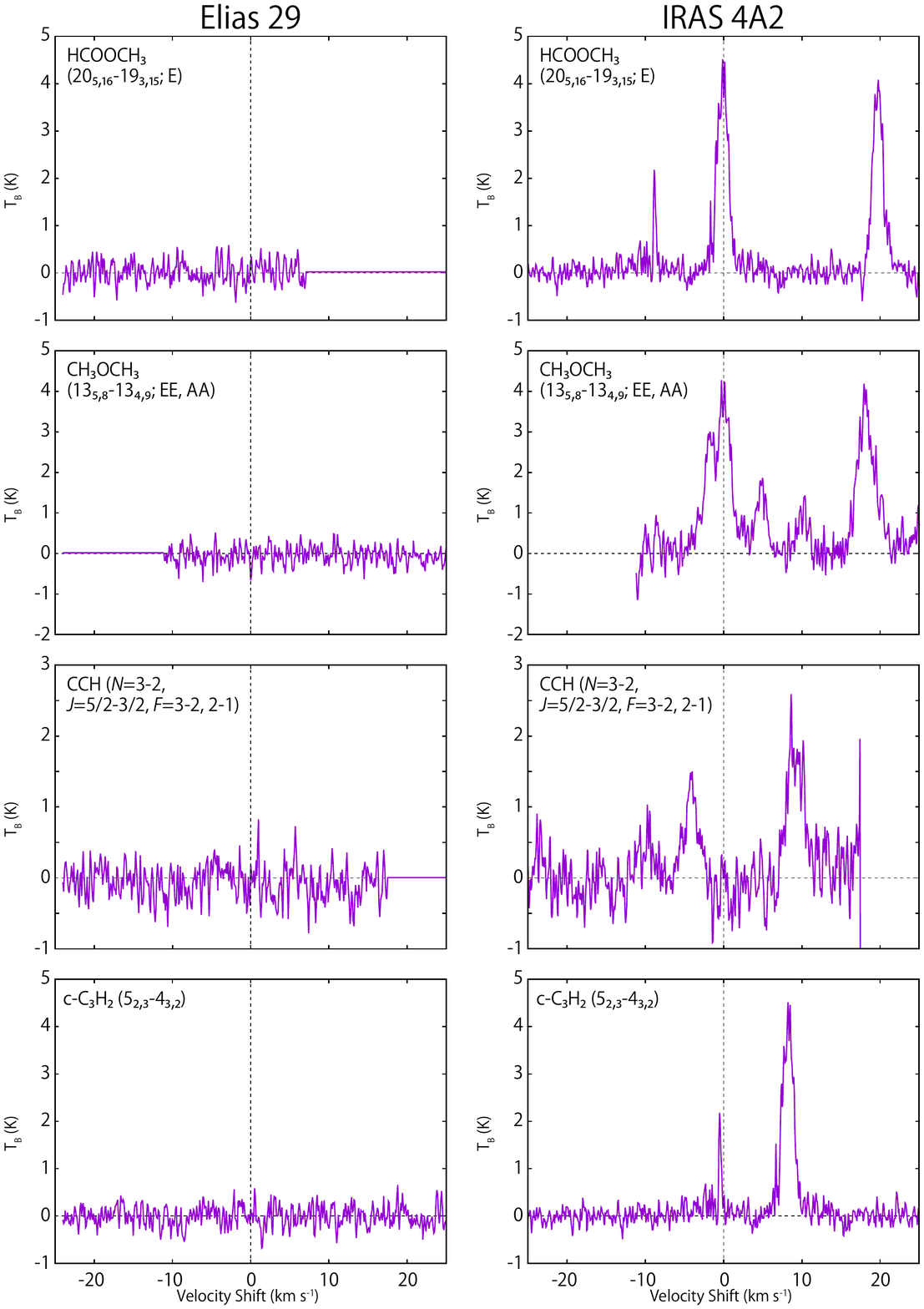}
	\vspace*{-17pt}
	\fi
	\caption{Spectra of the \MF, \DME, \cycCH, and CCH lines in Elias 29 and NGC1333 IRAS 4A2 \citep{LopezSepulcre2017}. 
			In Elias 29 (left panels), 
			the spectra {\bfresubfirst are} averaged {\bfresubfirst over} a circular region centered at the continuum peak
			with a diameter of {\bfresubfirst 0\farcs5, the} same as {\bfresubfirst in} Figure \ref{fig:spectra}. 
			These observations {\bfresubfirst were} performed as a single project 
			to investigate chemical diversity 
			{\bfresubsecond young} protostellar sources systematically, 
			and are {\bfresubfirst designed} to achieve almost {\bfresubfirst the} same sensitivity. 
			\label{fig:spectra_Cbearing}}
\end{figure}

In Figure \ref{fig:spectra}, 
the SO, \SOO, and SiO lines show a large velocity width over 20 \kmps. 
The 
{\bfresubsecond high-velocity} 
components likely come from the rotating gas in the vicinity of the protostar. 
The shapes of the spectral profiles are different from one another; 
the SO and SiO {\bfresubfirst emissions each have} a single peak near the \sysV\ (\vsys\ $\sim$ \vsysval\ \kmps), 
while the \SOO\ emission {\bfresubfirst is flatter} over a velocity shift of 5 \kmps. 

The spectral profile of the \SOO\ emission is asymmetric {\bfresubfirst with} {\bfresubsecond respect to} the \sysV\ (Figure \ref{fig:spectra}). 
More specifically, the red-shifted part is brighter than the blue-shifted {\bfresubfirst component.} 
This {\bfresubfirst is} confirmed in the PV diagrams (Figure \ref{fig:PV_34SO-SO2-SiO}), 
as discussed in Section \ref{sec:kinstr_pv}. 
{\bfresubfirst Two-dimensional Gaussian fitting} of the integrated intensity maps of {\bfresubfirst \SOO\ yields} blue- and red-shifted {\bfresubfirst peaks} 
{\bfresubfirst of} $329\pm10$ and $225 \pm 9$ m\Jypb\, with {\bfresubfirst integration over the} velocity-shift ranges 
$-4$ to $+4$ \kmps\ and $+4$ to $+12$ \kmps, {\bfresubfirst respectively,} 
the difference being a factor of 0.68. 

A similar asymmetry has been reported for the Class 0 low-mass protostellar source L483 {\bfresubfirst on a scale of} 100 au 
\citep[CS, SO, HNCO, \FA, \MF;][]{Oya_483}. 
It is suggested {\bfresubfirst that this is} due to the asymmetric distribution of these molecules, 
which {\bfresubsecond could be the case} 
for \SOO\ in Elias 29 as well. 
We note, however, that the line profiles of Elias 29 and L483 are {\bfresubfirst both red-shift deviated,} 
while the asymmetry of a gas distribution should be random, 
causing either {\bfresubfirst a red-shift or blue-shift deviated} profile {\bfresubfirst in principle.}  
In other words, {\bfresubfirst a red-shift deviated} line profile could 
{\bfresubfirst originate from a common physical reason relating to the vicinity of the protostar.} 
We {\bfresubfirst now} discuss {\bfresubfirst the following two} possibilities.

The weak blue-shifted emission could be explained 
if the dust in the vicinity of the protostar {\bfresubfirst were} optically thick at the corresponding frequency. 
We consider {\bfresubfirst an} edge-on configuration of the {\bfresubfirst \desys\ where} the molecular gas {\bfresubfirst is infalling.} 
Then, the blue-shifted emission of molecular lines from the back side could be 
{\bfresubsecond attenuated by dust.} 
The intensity of the blue-shifted emission {\bfresubfirst would} be attenuated by a factor of 
{\bfresubfirst 0.93, if we assumed an optical depth ($\tau$) of 0.07 for the dust (Section \ref{sec:dist_cont}).} 
This {\bfresubfirst attenuation could not} explain the observed difference of a factor of 0.68 seen in the \SOO\ emission. 
Here, we note that the above value for $\tau$ is an averaged value around the protostar, 
and hence, the optical depth {\bfresubfirst of the dust} would be higher {\bfresubfirst nearer to the protostar.} 
{\bfresubfirst For} a thin disk {\bfresubfirst with a constant density} 
with a radius of $r$ and a scale height of $d$, 
the optical depth averaged in the circle with a radius of $r$ 
is smaller than the actual optical depth {\bfresubfirst near the protostar} 
by a factor of $\dfrac{d}{2r}$. 
{\bfresubfirst This} is estimated from the volume between 
the thin disk and a cylinder with a radius of $r$ and a height of $2r$ 
surrounding the disk. 
If the dust continuum emission {\bfresubfirst came from a} thin-disk structure 
with a radius of 20 au and a height of 7 au, 
{\bfresubfirst $\tau$ would} be 0.4. 
If this were the case and there were infall motion in the vicinity of the protostar, 
the absorption of the blue-shifted emission by the dust would explain 
the observed intensity asymmetry in the \SOO\ line.

Alternatively, 
{\bfresubsecond the fact that the blue-shifted emission is weaker than the red-shifted emission} 
{\bfresubfirst could be} attributed to {\bfresubfirst expansion.} 
When the gas is expanding, 
the blue-shifted emission from the gas {\bfresubfirst in front {\bfresubsecond of the protostar} is} reduced 
in comparison with the red-shifted emission from the gas {\bfresubsecond to the rear of the protostar} 
\citep[e.g.,][]{Beals1953}. 
In this case, 
{\bfresubfirst reduction in intensity occurs mainly for a 
{\bfresubsecond low-velocity} region; 
this reduction is} due to the foreground molecular gas, 
which has an excitation temperature lower than {\bfresubfirst that of} the gas {\bfresubfirst near} the protostar. 
This {\bfresubsecond could be the case} in 
Elias 29 and L483 if the above molecular lines come from the gas in an outflow or a disk wind near the protostar. 
{\bfresubfirst However, we} note that the intensity of the 
{\bfresubsecond high-velocity} 
{\bfresubfirst region} ($v_{\rm shift} \sim -8$ \kmps) 
seems to be {\bfresubfirst reduced} in Elias 29, 
which is difficult to attribute to 
{\bfresubsecond gas in the foreground} 
of the expanding flow. 
{\bfresubfirst Nevertheless,} 
the possibility 
{\bfresubsecond that the asymmetry of the intensity} 
due to an outflow or disk wind cannot be excluded at this stage 
{\bfresubfirst and should be further tested.}


\section{Summary} \label{sec:summary}

We have analyzed 
ALMA Cycle 2 data {\bfresubsecond obtained for} 
various molecular lines (Table \ref{tb:lines}) 
{\bfresubfirst from the} Class I 
protostellar source Elias 29. 
{\bfresubfirst The major} findings are {\bfresubfirst summarized} below: 

\begin{enumerate}
\item[(1)] 
The SO and \SOO\ lines are bright in the compact region around the protostar 
{\bfresubsecond within a region of} 
diameter of a few 10s {\bfresubfirst of} au. 
The SO line also traces a ridge component 
{\bfresubfirst at a distance of 4\arcsec\ ($\sim$500 au) from the protostar toward the south.} 
SiO emission is detected {\bfresubfirst around the protostar.} 
{\bfresubfirst Meanwhile,} 
the CS emission is weak at the {\bfresubfirst protostar,} 
while it traces the southern ridge component. 
{\bfresubfirst Around the protostar,} 
the abundance ratio SO/CS is as high as $3^{+13}_{-2} \times 10^2$, 
which is 
even higher than that found in {\bfresubfirst an} outflow shocked region (L1157 B1). 

\item[(2)] 
Elias 29 is deficient 
in both saturated and unsaturated organic molecules, 
such as \MF, \DME, CCH, and \cycCH. 
Their deficiency as well as the richness in SO\ and \SOO\ can be explained qualitatively 
by chemical evolution {\bfresubfirst of a} Class I source or 
by the relatively high dust temperature (\gtsim 20 K) in the parent cloud of Elias 29 
in its prestellar core phase. 
{\bfresubfirst Determining which} of these two possibilities {\bfresubfirst applies here} is left for future study. 

\item[(3)] 
The SO and {\bfresubfirst \SOO\ emissions} show  
a velocity gradient along the north-south direction across the {\bfresubfirst protostar.} 
{\bfresubfirst Although the gradient} 
is likely due to {\bfresubfirst rotation} around the protostar, 
{\bfresubfirst it is difficult to distinguish between Keplerian motion and infall-rotation in our observation.} 
If we {\bfresubfirst assume} this 
{\bfresubsecond rotation} 
{\bfresubfirst to be} Keplerian for simplicity, 
the kinematic structure {\bfresubfirst that we} observed with the SO line can be reproduced by a protostellar mass 
{\bfresubfirst between 
{\bfresubsecond 0.82 and 1.0} 
\Msun\ {\bfresubsecond by assuming} 
an \ia\ of 90\degr\ to 65\degr} (\incRem). 


\item[(4)] 
{\bfresubfirst The \SOO\ spectrum is asymmetric, 
with} the blue-shifted components weaker than the red-shifted. 
Although {\bfresubfirst this asymmetry} can be attributed to an inhomogeneous molecular distribution, 
we need {\bfresubfirst to consider} other possible causes, 
such as dust opacity or {\bfresubfirst outflow motion.} 

\end{enumerate}

\acknowledgements
The authors are grateful to Ryohei Kawabe for invaluable discussion. 
{\bfresubfirst The authors also acknowledge the referee for a helpful 
{\bfresubsecond comments and suggestions} to improve {\bfresubsecond this} article.} 
This {\bfresubfirst study used} the ALMA data set ADS/JAO.ALMA\#2013.1.01102.S. 
ALMA is a partnership of the European Southern Observatory, the National Science Foundation (USA), the National Institutes of Natural Sciences (Japan), the National Research Council (Canada), and the NSC and ASIAA (Taiwan), 
in cooperation with the Republic of Chile. 
The Joint ALMA Observatory is operated by the ESO, the AUI/NRAO, and the NAOJ. 
The authors acknowledge the ALMA staff for their excellent support. 
This study is supported by a Grant-in-Aid from the Ministry of Education, Culture, Sports, Science, and Technologies of Japan 
(grant numbers 25400223, 25108005, 18H05222, {\bfresubsecond 19H05069, and 19K14753).} 
N.S. and S.Y. are grateful to financial support by JSPS and MAEE under the Japan-France Integrated Action Program (SAKURA: grant number 25765VC). 
C.C. and B.L. are grateful to financial support by Le Centre National de la Recherche Scientifique (CNRS) under the France-Japan action program.

\appendix

\section{Desorption Temperature} \label{sec:appendix_Tdes}

{\bfresubfirst The desorption} temperature of a molecular species (also called sublimation temperature) 
is {\bfresubfirst the} typical temperature at which the species thermally desorbs from dust grains. 
{\bfresubfirst According to \citet{Yamamoto_chembook},} 
the \desT\ (\Td) can be represented 
{\bfresubfirst in terms of the balance between desorption and adsorption} 
as: 
\begin{align}
	k_{\rm B} T_{\rm des} &= E_{\rm des} \left( \log \frac{\nu_0}{n_0 \Sigma <v>} \right)^{-1}, 
	\label{eq:app_desT}
\end{align}
{\bfresubfirst where 
\Ed\ denotes the \desE\ (or binding energy) of molecule X, 
$\nu_0$ the characteristic frequency of the vibration mode, 
$n_0$ the number density of H nuclei, 
$\Sigma$ the effective collision area of dust per H molecule, 
and $<v>$ the average speed of X. 
Typical values for 
$\nu_0$ and $\Sigma$ 
are $10^{12}$ Hz and $10^{-22}\ {\rm cm}^2$, {\bfresubfirst respectively} 
\citep{Hasegawa1992, Yamamoto_chembook}.} 

\begin{figure}[h]
	\begin{center}
	\iffigure
	\includegraphics[bb = 0 0 800 600, scale = 0.5]{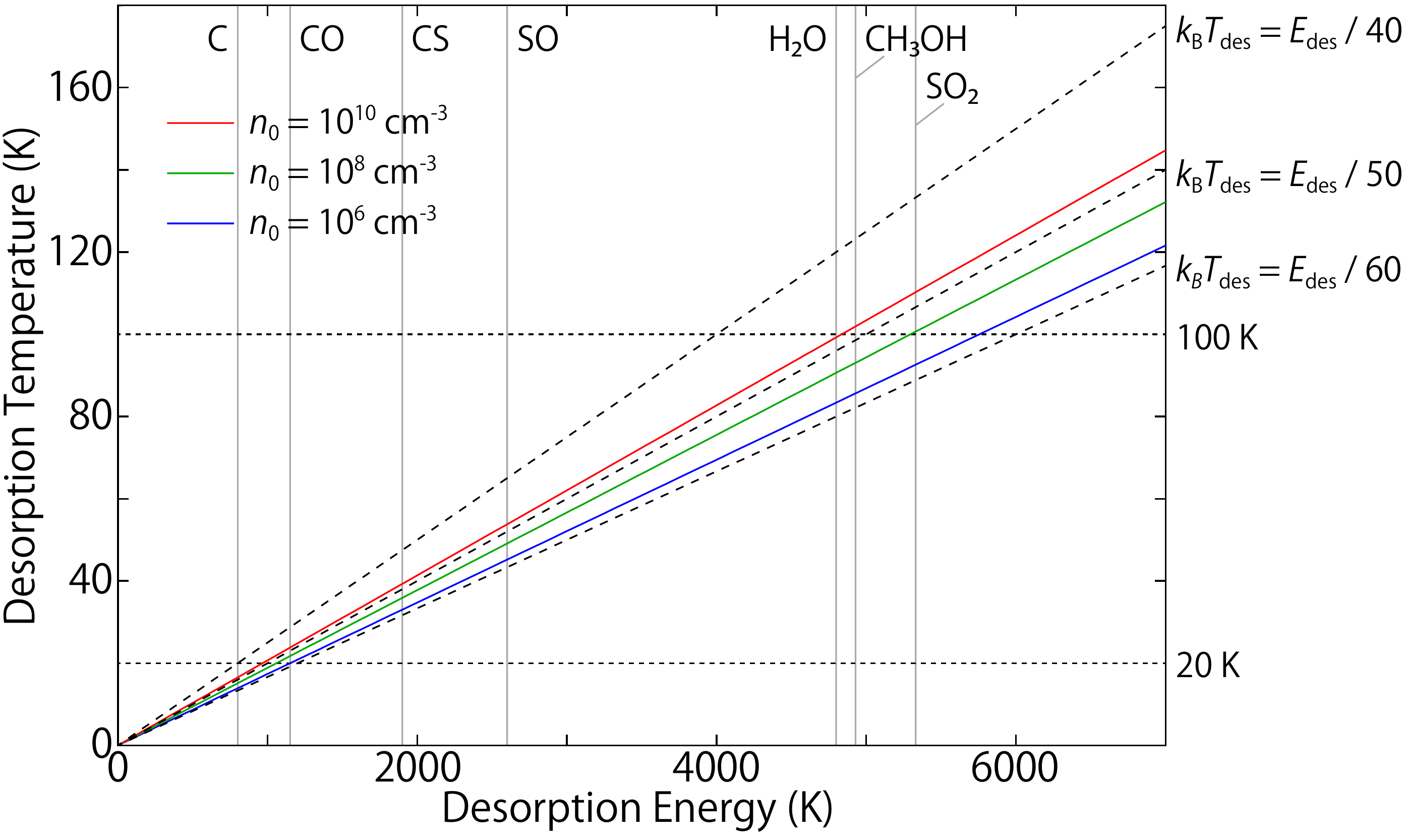}
	\fi
	\caption{Relation between the \desT\ and \desE. 
			{\bfresubfirst The solid} lines represent the plots derived from Equation (\ref{eq:app_desT}), 
			where $n_0$ and $<v>$ are assumed to be {\bfresubfirst $(10^6-10^{10})$ \cmcubic\ and $0.01$ \kmps, respectively. 
			The dashed lines are} plots of  
			$k_B T_{\rm des} = C E_{\rm des}$, 
			where $C$ denotes {\bfresubfirst a proportionality} factor of ($40-60$). 
			\label{fig:app_desTE}}
	\end{center}
\end{figure}

Here, the \desT\ of X is proportional to its \desE. 
Figure \ref{fig:app_desTE} shows the relation between the \desT\ and the \desE\ with typical values for $n_0$ and $<v>$. 
From the plots, the proportionality {\bfresubfirst factor} of \Td\ to \Ed\ is {\bfresubfirst approximately} ($50-60$), 
which weakly depends on $n_0$, $\nu_0$, $\Sigma$, and $<v>$. 
{\bfresubfirst Although} \Td\ tends to be higher for higher $n_0$, 
{\bfresubfirst this proportionality relation is practically useful for estimating \Td.} 
{\bfresubfirst By assuming} the factor to be 55 ($n_0$ of $10^8$ \cmsquare\ and $<v>$ of 0.1 \kmps), 
{\bfresubfirst we can estimate} the \desT s of {\bfresubfirst the} molecular species in this study as well as some representative molecular species, as summarized in Table \ref{tb:app_desT}.

\section{Protostellar Mass Estimation with Keplerian Motion and with Combined Infall and Rotation} 
\label{sec:appendix_mass}

In Keplerian motion, 
the {\bfresubfirst rotational} velocity (\vrot) of the gas around the protostar is represented as: 
\begin{equation}
	v_{\rm rot} (r) = \sqrt{\dfrac{GM}{r}}, \label{eq:vrot_Kep}
\end{equation}
where 
$M$ {\bfresubfirst denotes} the mass of the central protostar 
and $r$ the radial distance from the protostar. 
{\bfresubfirst On the other hand, 
the gas motion can be {\bfresubsecond a combination of infall and rotation.} 
If ballistic motion is assumed,} 
the rotational velocity (\vrot) and infall velocity (\vfall) are represented as\citep{Oya_15398}: 
\begin{align}
	v_{\rm rot} (r) &= \dfrac{1}{r} \sqrt{2 G M r_{\rm CB}}, \label{eq:vrot_IRE} \\ 
	v_{\rm fall} (r) &= \dfrac{1}{r} \sqrt{2 G M (r - r_{\rm CB})}, \label{eq:vfall_IRE}
\end{align}
{\bfresubfirst where} \rCB\ denotes the radius of the \cb, 
which is the perihelion for the infalling gas. 
At the \cb, 
the gas only {\bfresubfirst rotates,} without any radial motion. 

When we measure \vrot\ at a certain position 
{\bfresubfirst at a} distance $r$ from the protostar, 
we can estimate the protostellar mass ($M$) 
{\bfresubfirst to be $\displaystyle \frac{1}{G} r_{\rm CB} v_{\rm rot}^2$ 
by using equation (\ref{eq:vrot_Kep}) for Keplerian motion.
For 
{\bfresubsecond a combination of infall and rotation,} 
we need to specify \rcb. 
When we simply assume that the position} 
is the \cb, 
the protostellar mass is calculated to be $\dfrac{1}{2G} r_{\rm CB} v_{\rm rot}^2$ 
with Equation (\ref{eq:vrot_IRE}). 
This is half of the mass 
derived 
by assuming Keplerian motion. 


\begin{landscape}
\begin{table}
	\resettbnote
	\begin{center}
	\caption{Parameters of the Observed Lines\tablenotemark{a} 
			\label{tb:lines}}
	\steptbnote
	\begin{tabular}{llccccc}
		\hline \hline 
		Molecule & Transition & Frequency (GHz) & $E_{\rm u}$ (K) & $S\mu^2$ (Debye$^2$) & $A_{ij}$ ($s^{-1}$) & Synthesized Beam \\ \hline
		Continuum & (1.2 mm) & $244.91-264.26$ & & & & $0\farcs856 \times 0\farcs470$ (P.A. 95\fdegr17) \\ 
		CS & \cs & 244.9355565 & 35.3 & 19.2 & $2.98 \times 10^{-4}$ & $0\farcs884 \times 0\farcs522$ (P.A. 94\fdegr60) \\ 
		SO & \so & 261.8437210 & 47.6 & 16.4 & $2.28 \times 10^{-4}$ & $0\farcs832 \times 0\farcs488$  (P.A. 94\fdegr17) \\ 
		\SO & \tfso & 246.6634700 & 49.9 & 11.4 & $1.81 \times 10^{-4}$ & $0\farcs883 \times 0\farcs517$ (P.A. 94\fdegr64) \\ 
		\SOO & \soo & 245.5634219 & 72.7 & 14.5 & $1.19 \times 10^{-4}$ & $0\farcs885 \times 0\farcs520$ (P.A. 94\fdegr78) \\ 
		SiO & \sio & 260.5180090 & 43.8 & 57.6 & $9.12 \times 10^{-4}$ & $0\farcs835 \times 0\farcs488$ (P.A. 94\fdegr67) \\ 
		\hline 
	\end{tabular}
	\end{center}
	\resettbnote
	\tbnotetext{Taken from CDMS \citep{Muller_CDMS}. } 
\end{table}
\end{landscape}

\begin{landscape}
\begin{table}
	\resettbnote
	\begin{center}
	\caption{1.2 mm Dust Continuum Parameters 
			\label{tb:continuum}}
	\begin{tabular}{ccccc}
	\hline \hline 
	Peak Intensity\tbnotemark & Integrated Flux\tablenotemark{a} & Assumed & $M_{\rm gas}$\tbnotemark & \nhydro \tablenotemark{b}  \\ 
	$I (\nu)$ (m\Jypb) & $F (\nu)$ (mJy) & Dust Temperature & ($10^{-3}$ \Msun) & ($10^{23}$ \cmsquare) \\
	\hline 
	$17.2 \pm 0.3$ & $21.4 \pm 0.7$ & 50 K & $1.68 \pm 0.06$ & $3.7 \pm 0.4$ \\ 
	& & 100 K & $0.79 \pm 0.03$ & $1.7 \pm 0.2$ \\ 
	& & 150 K & $0.52 \pm 0.02$ & $1.1 \pm 0.1$ \\ 
	\hline
	\end{tabular}
	\end{center}
	\resettbnote
	\tbnotetext{The peak intensity and integrated flux are obtained from two-dimensional Gaussian fitting.}
	\tbnotetext{The gas mass ($M_{\rm gas}$) and column density of \hydro\ (\nhydro) are evaluated 
			assuming the mass absorption coefficient to the dust mass ($\kappa_\nu$) of 1.3 \cmsquare g\inv\ at a wavelength of 1.2 mm 
			\citep{Ossenkopf1994}. 
			The errors are taken to be 3$\sigma$ for the integrated flux of the continuum emission.}
\end{table}
\end{landscape}

\begin{table}
	\resettbnote
	\begin{center}
	\caption{Observed Intensities and Derived Abundances at the Continuum Peak
			\label{tb:abundance}}
	\begin{tabular}{lcccc}
	\hline \hline 
	\multirow{2}{*}{Species} & $I (\nu)$\tbnotemark & Assumed & Column Density\tbnotemark & Fractional Abundance\tbnotemark \\
	& (\Jypb\ \kmps) & Gas Temperature & $N (X)$ (\cmsquare) & $f (X)$ \\ 
	\hline
	CS & $0.116\pm0.03$ & 50 K & $(1.6\pm1.2) \times 10^{13}$ & $(4.2\pm3.3) \times 10^{-11}$ \\ 
	& & 100 K & $(2.2\pm1.7) \times 10^{13}$ & $(1.3\pm1.0) \times 10^{-10}$ \\ 
	& & 150 K & $(2.9\pm2.3) \times 10^{13}$ & $(2.6\pm2.0) \times 10^{-10}$ \\ 
	SO\tbnotemark & $2.83\pm0.33$ & 50 K & $(4.6\pm1.4) \times 10^{15}$ & $(1.3\pm0.4) \times 10^{-8}$ \\
	& & 100 K & $(6.0\pm1.8) \times 10^{15}$ & $(3.5\pm1.0) \times 10^{-8}$ \\ 
	& & 150 K  & $(7.8\pm2.3) \times 10^{15}$ & $(6.9\pm2.0) \times 10^{-8}$ \\ 
	\SO & $0.225\pm0.0966$ & 50 K & $(2.0\pm0.6) \times 10^{14}$ & $(5.6\pm1.6) \times 10^{-10}$ \\
	& & 100 K & $(2.7\pm0.8) \times 10^{14}$ & $(1.5\pm0.5) \times 10^{-9}$ \\ 
	& & 150 K & $(3.5\pm1.0) \times 10^{14}$ & $(3.1\pm0.9) \times 10^{-9}$ \\ 
	\SOO\ & $0.621\pm0.052$ & 50 K & $(2.2\pm0.6) \times 10^{15}$ & $(6.0\pm1.5) \times 10^{-9}$ \\
	& & 100 K & $(3.0\pm0.8) \times 10^{15}$ & $(1.7\pm0.4) \times 10^{-8}$ \\ 
	& & 150 K & $(4.4\pm1.1) \times 10^{15}$ & $(3.9\pm1.0) \times 10^{-8}$ \\ 
	SiO & $0.113\pm0.024$ & 50 K & $(6.4\pm4.1) \times 10^{12}$ & $(1.7\pm1.1) \times 10^{-11}$ \\
	& & 100 K & $(8.3\pm5.3) \times 10^{12}$ & $(4.8\pm3.0) \times 10^{-11}$ \\ 
	& & 150 K & $(1.1\pm0.7) \times 10^{13}$ & $(9.5\pm6.0) \times 10^{-11}$\\ 
	\hline
	\end{tabular}
	\end{center}
	\resettbnote
	\tbnotetext{The peak integrated intensities are derived by using two-dimensional Gaussian fitting. 
			The errors represent 3$\sigma$ in the fitting. } 
	\tbnotetext{The column densities are derived from the integrated intensities 
			assuming LTE with the gas temperature ranging from 50 to 150 K. } 
	\tbnotetext{Fractional abundances relative to \hydro\ are calculated. 
			The column density of \hydro\ ($N$(\hydro)) is derived from the 1.2 mm continuum emission (Table \ref{tb:continuum}). 
			The gas and dust temperatures are assumed to {\bfresubfirst be} equal. }
	\tbnotetext{The column density and fractional abundance of SO {\bfresubfirst are determined} from {\bfresubfirst those} of \SO\ assuming the \SO/SO ratio of 22.6. } 
\end{table}

\begin{landscape}
\begin{table}
	\resettbnote
	\begin{center}
	\caption{Column Densities and Abundance Ratios of Sulfur-Bearing Species Compared with Other Sources 
			\label{tb:comparison}}
	\setcounter{tbnotecount}{2}
	\begin{tabular}{lccccc}
		\hline \hline 
		& Elias 29 & Elias 29 & \multirow{2}{*}{NGC 1333 IRAS 2\tbnotemark}& \iras \tbnotemark & \multirow{2}{*}{L1157 B1\tbnotemark} \\ 
		& Continuum Peak\tablenotemark{a} & Ridge\tablenotemark{b} & & Source B & \\ \hline
		$N$(CS) /\cmsquare & $(2.2\pm1.7) \times 10^{13}$ & $(5.1\pm1.4) \times 10^{13}$ & $(0.5-4.5) \times 10^{13}$& $(4.4-66) \times 10^{15}$ & $2.7 \times 10^{14}$ \\
		$N$(SO) /\cmsquare & $(6.0\pm1.8) \times 10^{15}$ & $(9.8\pm2.6) \times 10^{14}$ & $(0.6-3.5) \times 10^{14}$ & $< 5.0 \times 10^{14}$ & $(3-5) \times 10^{14}$ \\
		$N$(\SOO) /\cmsquare & $(3.0\pm0.8) \times 10^{15}$ & $(1.3\pm1.2) \times 10^{15}$ & $(0.5-7.0) \times 10^{13}$ & $1.5 \times 10^{15}$ & $3.0 \times 10^{14}$ \\ 
		\hline 
		\molratio{SO}{CS} & $3^{+13}_{-2} \times 10^2$ & $19^{+14}_{-8}$ & $1.7-37.5$ &  $< 0.11$ & $1.1-1.9$ \\ 
		\molratio{\SOO}{CS} & $1.4^{+6}_{-0.8} \times 10^2$ & $25^{+43}_{-24}$ & $0.1-6.3$ &  $0.02-0.34$ & $1.1$ \\ 
		\molratio{\SOO}{SO} & $0.5^{+0.4}_{-0.2}$ & $1.3^{+2.2}_{-1.3}$ & $0.05-0.3$ &  $> 3$ & $0.6-1.0$ \\ 
		\hline 
	\end{tabular}
	\end{center}
	\vspace*{-20pt}
	\resettbnote
	\tbnotetext{Molecular abundances in Elias 29 at the continuum peak 
				are derived {\bfresubfirst assuming} LTE and {\bfresubfirst a} gas temperature of 100 K \citep{RochaPilling2018}. 
				Errors are {\bfresubfirst taken to be} three times the rms noise of the integrated intensity. 
				The column density of SO is derived from the \SO\ emission 
				by assuming $^{32}$SO/\SO\ = 22.6. } 
	\tbnotetext{Molecular abundances in the ridge of Elias 29 
				are derived from the peak of the beam-averaged intensity 
				{\bfresubfirst assuming} LTE and {\bfresubfirst a} gas temperature of 20 K \citep{RochaPilling2018}. 				
				The peak intensity positions are:  
				$(\alpha_{2000}, \delta_{2000}) = (16^{\rm h}27^{\rm m}09\fs499, -24\degr37^\prime23\farcs942)$, 
				$(16^{\rm h}27^{\rm m}09\fs499, -24\degr37^\prime23\farcs511)$, 
				and $(16^{\rm h}27^{\rm m}09\fs490, -24\degr37^\prime23\farcs425)$ 
				for the CS, SO, and \SOO\ lines, respectively. 
				Errors are {\bfresubfirst taken to be} three times the rms noise of the integrated intensity. } 
	\tbnotetext{Taken from \citet{Wakelam2005}. 
				The values show ranges for the {\bfresubfirst 15} positions consisting of the protostellar position and positions in the outflow lobes. 
				} 
	\tbnotetext{Taken from \citet{Drozdovskaya_PILS}. } 
	\tbnotetext{A shocked region taken from \citet{Bachiller_L1157}. }
	\resettbnote
\end{table}
\end{landscape}

\begin{landscape}
\begin{table}
	\resettbnote
	\begin{center}
	\caption{Abundances of Organic Molecules at the Continuum Peak
			\label{tb:abundance_Cbearing}}
	\begin{tabular}{cccccc}
	\hline \hline 
	\multirow{2}{*}{Species} & \multicolumn{3}{c}{Elias 29} & NGC1333 IRAS 4A1\tbnotemark & NGC1333 IRAS 4A2\tablenotemark{a} \\ 
	& Integrated Intensity\tbnotemark & Column Density\tbnotemark & \molratio{$X$}{\hydro}\tbnotemark & \molratio{$X$}{\hydro}\tablenotemark{d} & \molratio{$X$}{\hydro}\tablenotemark{d} \\ 
	\hline 
	\MF & $<5$ & $<(1.8-2.3) \times 10^{14}$ & $<(0.6-2.0) \times 10^{-9}$ & $<8.2 \times 10^{-11}$ & $(1.1 \pm 0.1) \times 10^{-8}$ \\ 
	\DME & $<5$ & $<(2.9-3.6) \times 10^{14}$ & $<(0.9-3.2) \times 10^{-9}$ & $<4.3 \times 10^{-10}$ & $(1.0 \pm 0.1) \times 10^{-8}$ \\ 
	CCH & $<7$ & $<(4.7-10.1) \times 10^{12}$ & $<(1.3-9.0) \times 10^{-11}$ & & \\ 
	\cycCH & $<9$ & $<(3.5-10.4) \times 10^{12}$ & $<(0.9-9.2) \times 10^{-11}$ & & \\ 
	\hline 
	\end{tabular}
	\end{center}
	\resettbnote
	\tbnotetext{Fractional abundances relative to \hydro\ in NGC1333 IRAS 4A1 and 4A2 are taken from \citet{LopezSepulcre2017}.} 
	\tbnotetext{In m\Jypb. The rms noise levels (1$\sigma$) of the integrated intensity maps 
			(Figure \ref{fig:mom0_Cbearing}) are employed as the upper limits.} 
	\tbnotetext{In \cmsquare. Upper limits to the column densities are derived from those to the integrated intensities assuming LTE 
			with the gas temperature ranging from 50 to 150 K.} 
	\tbnotetext{Upper limits to the fractional abundances relative to \hydro\ are calculated. 
			The column density of \hydro\ (\nhydro) is derived from the 1.2 mm continuum emission (Table \ref{tb:continuum}). 
			The gas and dust temperatures are assumed to be equal.} 
\end{table}
\end{landscape}

\setcounter{tbnotecount}{0}
\begin{table}
	\begin{center}
	\caption{Desorption Energies and Desorption Temperatures of Molecular Species \label{tb:app_desT}}
	\begin{tabular}{ccc}
	\hline \hline 
	\multirow{2}{*}{Molecular Species} & Desorption Energy\tbnotemark & Desorption Temperature\tbnotemark \\ 
	& \Ed$/k_B$ (K) & \Td (K) \\ \hline 
	C & 800 & 15 \\ 
	S & 1100 & 20 \\ 
	CO & 1150 & 21 \\ 
	O & 1600 & 30 \\ 
	CS & 1900 & 35 \\ 
	\FAD & 2050 & 37 \\ 
	CCH & 2137 & 39 \\
	SO & 2600 & 47 \\ 
	\TFA & 2700 & 49 \\ 
	H$_2$S & 2743 & 50 \\ 
	OCS & 2888 & 53 \\ 	
	SiO & 3500 & 64 \\ 
	\MF & 4000 & 73 \\ 
	H$_2$O & 4800 & 87 \\ 
	\MN & 4930 & 90 \\ 
	HCOOH & 5000 & 91 \\ 
	SO$_2$ & 5330 & 97 \\ 
	\FA & 5556 & 101 \\ 
	\hline 
	\end{tabular}
	\end{center}
	\setcounter{tbnotecount}{0}
	\tbnotetext{Taken from Kinetic Database for Astrochemistry 
			\citep[KIDA; ][\url{http://kida.obs.u-bordeaux1.fr/}]{Wakelam_KIDA}} 
	\tbnotetext{Derived from the desorption energy with the equation 
			$k_{\rm B} T_{\rm des} = E_{\rm des} / 55$. 
			This simplified relation is obtained using {\bfresubfirst Equation (\ref{eq:app_desT}), 
			assuming $n_0 = 10^7$ \cmcubic\ and $< v > =$ 0.01 \kmps.} 
			}
\end{table}

\clearpage

\clearpage

\clearpage

\clearpage

\clearpage

\clearpage

\end{document}